\documentclass{aa}  
\usepackage{graphicx}
\usepackage{txfonts}
\usepackage{booktabs}
\usepackage{gensymb}
\usepackage{hyperref}
\usepackage{multirow}
\usepackage{xcolor}
\usepackage[normalem]{ulem}
\usepackage{CJKutf8}

\newcommand{\CNnames}[1]{{\begin{CJK}{UTF8}{gbsn}~(#1)~\end{CJK}}}

\begin{document} 
   \title{Isochrone-cloud fitting and asteroseismology of the \emph{Kepler} open cluster NGC\,6866}
%   \subtitle{}

   \author{Haotian Wang\CNnames{王昊天}
          \inst{1}
          \and Gang Li\CNnames{李刚}\inst{1,2}\and Dario J. Fritzewski\inst{1}\and Timothy Van Reeth\inst{1}\and Conny Aerts\inst{1,3,4}\fnmsep
          }

   \institute{
             Institute of Astronomy, KU Leuven, Celestijnenlaan 200D, 3001 Leuven, Belgium\\
             \email{haotian.wang@kuleuven.be}
             \and Centre for Astrophysics, University of Southern Queensland, Toowoomba,
QLD 4350, Australia
            \and Department of Astrophysics, IMAPP, Radboud University Nijmegen, PO Box 9010, 6500 GL Nijmegen, The Netherlands
            \and Max Planck Institut für Astronomie, Königstuhl 17, 69117 Heidelberg, Germany\\
             }

   \date{}

\abstract
  % context heading (optional)
   {Isochrone fitting is a classical method used to determine the ages of open clusters. Yet the derived ages depend strongly on the chosen input physics of the stellar models used to deduce the isochrones. Additionally, it remains an open question whether the isochronal ages based on stellar surface parameters are consistent with the asteroseismic ages derived from stellar interior properties.}
  % aims heading (mandatory)
   {We aim to investigate how isochrones based on different input physics and computed for a variety of initial conditions affect age dating of the open cluster NGC\,6866. By using the 4-year-long light curves of its members assembled by the \emph{Kepler} mission, we further compare these results with asteroseismically derived ages.}
  % methods heading (mandatory)
   {We extracted 180 cluster members by applying a clustering algorithm to Gaia Data Release 3 (DR3) data, and subsequently developed an “isochrone-cloud” fitting method that simultaneously accounts for the range of free parameters occurring in the input physics. Synthetic colour–magnitude diagrams (CMDs) with various initial rotation rates were generated and compared with the observations to determine the best-fitting rotation distribution. Variable stars were then identified among the cluster members. For the 19 gravity-mode (g-mode) pulsators, we performed modelling using a dedicated grid of rotating stellar models. This modelling was constrained by stellar surface parameters, the measured asymptotic gravity-mode (g-mode) period spacing $\Pi_0$, and the near-core rotation rate. Two approaches were considered: modelling the pulsators individually and under the assumption that they share a common age.}
  % results heading (mandatory)
   {We found discrepant age estimates from PARSEC and MIST isochrones, which yielded $690^{+140}_{-30}\,\mathrm{Myr}$ and $467^{+70}_{-50}\,\mathrm{Myr}$, respectively. The isochrone-cloud fitting indicates that NGC\,6866 has an initial rotation distribution peaking at $v/v_\mathrm{crit}=0.6$, which is a discrepancy with a factor of about two compared to the asteroseismic values. The asteroseismic modelling of the g-mode pulsators revealed agreement between seismic and isochronal masses, whereas the derived ages differ substantially due to the difference in internal mixing. When the g-mode pulsators were modelled under the assumption of one shared cluster age, we obtained a value of $759^{+54}_{-82}\,\mathrm{Myr}$, in agreement with the PARSEC-based isochronal age.}
  % conclusions heading (optional)
   {We conclude that using different input physics and various initial conditions impacts the age-dating results of open clusters. Our findings point to the need for more carefully calibrated evolutionary models. The seismic content of NGC\,6866 makes it a suitable middle-aged cluster to achieve such calibrations.}


   \maketitle
%
%-------------------------------------------------------------------
\nolinenumbers
\section{Introduction}

   Open clusters are indispensable benchmark objects for stellar astrophysics, in particular for calibrating stellar evolution models. This stems from the fact that stars in open clusters are approximately coeval and chemically homogeneous \citep{lada2003oc}. Yet, turning this advantage into precise ages and interior properties remains challenging: colour–magnitude diagram (CMD) fits are sensitive to the adopted input physics--convective-core overshoot, rotation and angular-momentum transport, envelope mixing, and extinction, making different model families yield different answers \citep{choi2016,bressan2012}.
   
   Since member stars in open clusters are expected to share the same age and chemical composition, those without stellar companions should ideally follow a single isochrone track in the CMD, with variations only due to their initial masses \citep{choi2016}. Yet, observations of the main-sequence turn-off (MSTO) region--where stars evolve from the main sequence towards the red giant branch, often show a broadening that cannot be explained by an isochrone with a single set of input physics and fixed values for its free parameters, especially in young and intermediate-age clusters. This phenomenon, known as the extended main-sequence turn-off (eMSTO; \citealp{bastian2009, LiCY2024}), challenges the classical view of a single isochrone per cluster.
   
   Several explanations have been proposed for the eMSTO, including multiple stellar populations \citep{2007Mackey}, stellar rotation \citep{2018groudfrooij} and binarity \citep{2022he}. However, none of these individual explanations can fully account for the observed morphology. In particular, they cannot reproduce both the observed concentration of stars at the bottom of the red giant branch and the eMSTO \citep{2018bastian,2019gossage}.

   Isochrone fitting, despite its limitations in accounting for the eMSTO phenomenon, remains one of the most widely used, consistent and successful methods for determining the ages of stellar populations in open clusters \citep{2001yi, LiCY2024}. Conventionally, this is done by fitting a single isochrone to an observed cluster CMD. However, this approach has clear drawbacks: each isochrone represents a unique set of input physics with free parameters -- including age, extinction, convective core overshoot, and a single value of the initial rotation rate. This limits the method for modelling populations with a diversity of parameters, especially for the core overshoot and the initial rotation \citep{2019johnston,2025reyes}.

   One of the approaches to address this issue is to generate a synthetic stellar population with a variety of initial physical properties and compare it with observations. In globular clusters, comparing the synthetic CMD with the observed CMD through Voronoi binning has been proven effective \citep{2023ying}. However, this comparing technique is less applicable to open clusters, due to their sparser stellar populations and less densely populated CMDs. Therefore, a suitable methodology for open clusters is required—one that can more robustly address age determination, incorporate variations in stellar input physics, and integrate asteroseismic constraints on individual stars within the population. To provide the required, independent constraints on internal structure and rotation, a CMD-based population modelling with star-by-star asteroseismic diagnostics on the internal physics is required.

   Asteroseismic constraints arise from stellar oscillations and are especially powerful in A–F main-sequence stars. In the range $7,000$–$10,000$\,K, classical pulsations are commonly observed and
   typically classified as either $\delta$\,Scuti ($\delta$\,Sct) stars exhibiting 
   pressure (p) modes \citep{1979breger,breger1993,2000breger,2011uytterhoeven}
or $\gamma$\,Doradus ($\gamma$\,Dor) stars with gravity (g) modes \citep{Kaye1999,2011uytterhoeven}. 
The high-order g modes in $\gamma$\,Dor stars are sensitive to conditions near the convective core, enabling constraints on core mass and near-core rotation. Because these core properties evolve with time, they provide age-sensitive diagnostics that directly complement CMD-based estimates \citep[e.g.:][]{vanreeth2016, 2019aerts, ouazzani2019, li2020, mombarg2021,  Mombarg2023}.

   Observationally, the amplitude spectrum of main-sequence g-mode pulsators often displays long series of oscillation frequencies, representing modes of consecutive radial order $n$ for fixed spherical-harmonic degree $\ell$ and azimuthal order $m$. The mode period differences between consecutive modes of the same $(\ell,m)$, $\Delta P \equiv P_{n+1}-P_n$, form a characteristic $\Delta P$–$P$ pattern, called the period-spacing pattern. The traditional approximation of rotation (TAR) provides a framework for modelling this period-spacing pattern in main-sequence pulsators \citep{eckart1960, 1987lee,lee1997, 2003townsend, vanreeth2016}.  As described in \cite{vanreeth2016}, in the asymptotic limit under the TAR, $\Delta P \simeq \Pi_0/\sqrt{\lambda_{\ell,m,s}}$, where $\Pi_0$ is the buoyancy radius and $\lambda_{\ell,m,s}$ is the eigenvalue of Laplace’s tidal equation that depends on $\ell, m$ and the spin parameter $s$ defined as twice the rotation frequency divided by the oscillation frequency in a co-rotating frame of reference. Rotation thus imprints a systematic trend: prograde sequences $(m>0)$ typically show a decreasing $\Delta P$ with $P$, retrograde sequences $(m<0)$ an increasing trend, and zonal sequences remain comparatively flat \citep{Bouabid2013,VanReeth2015-method}. By fitting the mean level of the pattern (constraining $\Pi_0$) and its slope (constraining the near-core rotation rate), one can obtain joint constraints on internal structure \citep[e.g.][]{vanreeth2016,Michielsen2021}, envelope mixing \citep[e.g.][]{2021pedersen}, and asteroseismic ages \citep{fritzewski2024}.
   
   In the past decade, space missions like \emph{Gaia} \citep{2016gaiacollab} and \emph{Kepler} \citep{borucki2010} have revolutionised stellar astrophysics by providing precise measurements of astrometric and photometric properties. These high-precision datasets have opened new avenues for studying open clusters with unprecedented detail \citep{2018cantat-gaudin, babusiaux2023}. The \emph{Kepler} space telescope, in particular, continuously monitored a field near the Galactic plane for four years, obtaining high-quality photometry for hundreds of thousands of stars. This long-term, uninterrupted dataset had a transformative impact on asteroseismology \citep{2021aerts}, enabling detailed studies of stellar interiors across a broad range of variable stars \citep[e.g.][]{Bedding2011Natur, Stellor2016Natur, Bedding2020, 2021pedersen, Li2022, LiYaGuang2022NatAs, Reyes2025Natur}. 
   
   Here, we present joint open cluster and asteroseismic modelling of NGC\,6866 which was observed by the {\it Kepler\/} mission, to constrain the influence of the assumed input physics on stellar populations. This paper is organised as follows. In Sect.~\ref{sec:target}, we introduce the target open cluster NGC\,6866, including its basic properties and a summary of previous studies from the literature along with our membership identification. Sect.~\ref{sec:icloud} presents our customised isochrone-cloud fitting for NGC\,6866. In Sect.~\ref{sec:variability}, we classify the variable stars among the identified member stars using \emph{Kepler} photometry, and we provide a catalogue of the variable stars. In Sect.~\ref{sec:asteroseismology}, we focus on forward modelling of the detected g-mode pulsators, comparing their asteroseismic masses and ages with those inferred from isochrone fitting. Finally, we give our conclusions in Sect.~\ref{sec:conclusion}.

\section{Target Cluster NGC\,6866}
\label{sec:target}
\subsection{Previous studies on NGC\,6866}
   Originally discovered by Caroline Herschel \citep{herschel1802}, the open cluster NGC\,6866 ($\alpha = 300\degree.9792$, $\delta = +44\degree.1583$; \citealp{Cantat-Gaudin2020}) is a bright, relatively young cluster located in the constellation Cygnus. It has a mean distance of $1406$~pc, based on the parallax data from \emph{Gaia} DR3 \citep{poggio2021gedr3}, and a metallicity consistent with the solar value \citep{bostanc2015}.
   
   Before the launch of the \emph{Kepler} spacecraft, studies of NGC\,6866 primarily focused on its Galactic kinematics and chemical abundances \citep[e.g.][]{1961johnson, 1971becker}. The availability of high-precision photometry from the \emph{Kepler} mission has since enabled more detailed investigations of the properties of individual stars, including stellar variabilities due to pulsations and rotations \citep{Balona2013}. Age estimates for NGC\,6866 vary widely, ranging from 430 to 780\,Myr, including 430\,Myr derived from red giant asteroseismology \citep{brogaard2023}, 480\,Myr from main-sequence isochrone fitting \citep{Kharchenko2005}, 650\,Myr from an artificial neural network analysis applied to Gaia DR2 data \citep{Cantat-Gaudin2020}, and 780\,Myr from traditional isochrone fitting \citep{bossini2019}.

   This spread in age estimates arises partly from differences in the methodologies used to match theoretical stellar models to observed CMDs. Discrepancies also stem from the choice of input physics in the stellar evolution models used, such as PARSEC \citep{bressan2012}, MIST \citep{dotter2016, choi2016}, or Geneva \citep{2001lejeune}. For example, \citet{brogaard2023} noted that PARSEC models tend to yield systematically older ages, exceeding the MIST-based values by more than 1\,$\sigma$.

   Variations in input physics, such as the treatment of convective core overshooting and internal stellar rotation, further complicate the age determination. On the observational side, the CMD of NGC\,6866 exhibits an underpopulated eMSTO region and lacks subgiants, limiting the ability to constrain the isochrone fit. Additionally, most current isochrone-fitting methods struggle to quantitatively assess the residuals of stars deviating from the best-fit model, particularly near the base of the red giant branch, where theoretical isochrones often dip before ascending \citep{brogaard2023}.

   Since NGC\,6866 is located within the \emph{Kepler} field, it has been extensively monitored by spectroscopic sky surveys such as LAMOST \citep{cui2012RAA, decat2015}. Approximately half of the previously identified cluster members have atmospheric chemical abundance measurements. \citet{bostanc2015} utilised low-resolution LAMOST spectra for stars within a 6-arcminute radius and derived a metallicity of $\mathrm{[Fe/H]}=-0.1$ dex, corresponding to $Z = 0.01$. They concluded that adopting a solar metallicity is a reasonable approximation for this cluster.

   The majority of NGC\,6866's member stars were observed by the \emph{Kepler} spacecraft, which provided four years of continuous, high-precision photometry suitable for variability studies. \citet{brogaard2023} analysed red giant oscillations in the cluster using these \emph{Kepler} light curves. By applying a modified scaling relation to five helium-burning giants, they estimated precise stellar masses for these evolved stars, with a mean stellar mass of approximately $2.8\,\mathrm{M_\odot}$. This implies that the cluster's main sequence may contain pulsators of intermediate mass, such as $\gamma$\,Dor and $\delta$\,Sct stars. Such pulsators are a main focus of our work.

\subsection{Cluster membership identification}

   We performed the cluster membership identification for NGC\,6866 using a Gaussian Mixture Model (GMM) clustering algorithm \citep{mclachlan2019finite}, which estimates the probability of each data point belonging to a specific component of a mixture of Gaussian distributions. The input sample consisted of all stars with available astrometric measurements in \emph{Gaia} DR3 within an angular radius of 56 arcminutes from the cluster's central coordinates. This radius corresponds to four times the tidal radius ($14.4$ arcminutes), as reported by \citet{nilakshi2002} and \citet{drury2020}. The measurements used for the clustering are the equatorial coordinates ($\alpha$; $\delta$), proper motions in right ascension and declination ($\mu_{\alpha,\star}$, $\mu_{\delta,\star}$), and parallax ($\varpi$), along with their respective uncertainties, obtained from \emph{Gaia} DR3 \citep{2021gaiacollab, 2023gaiacollab}. A model with 14 Gaussian components for the GMM is adopted for the fit. The number was selected based on the local minimum of the Bayesian Information Criterion (BIC) evaluated for different numbers of Gaussian components.

   After determining the optimal number of kernels, we excluded stars with $m_G > 16$\,mag since stars with $m_\mathrm{Kepler} > 16$\,mag are less likely to have \emph{Kepler} light curve data. Additionally, applying a magnitude cut also cuts off stars with high uncertainties in their astrometric measurements. To assess the uncertainty in the clustering, we performed a perturbative Monte Carlo process. In each iteration, random perturbations were applied to the astrometric parameters $\alpha, \delta, \mu_{\alpha}, \mu_{\delta}$, and $\varpi$. The perturbations were drawn from a normal distribution with a mean equal to the measured value and a standard deviation equal to the associated uncertainty.

\begin{figure*}
        \includegraphics[width=\linewidth]{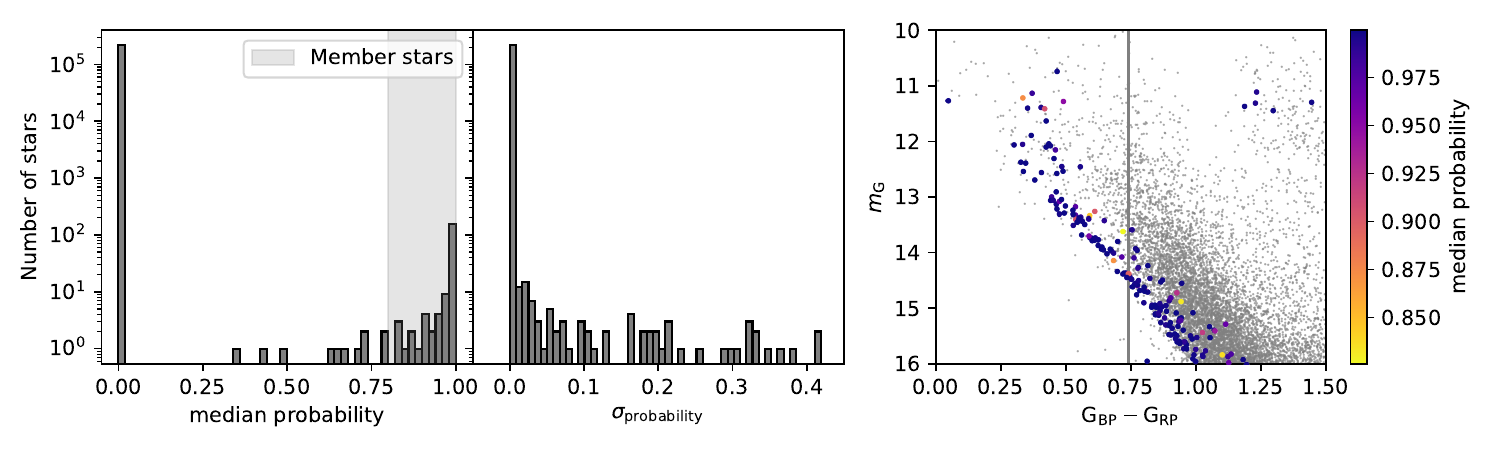}
    \caption{Clustering algorithm results for membership identification. The left and middle panels show the median and standard deviation of membership probabilities for all stars within the radius cut. The shaded region indicates the selected members with $p>0.8$. The right panel displays the colour-magnitude diagram (CMD) of NGC\,6866, where the coloured markers represent the 180 identified member stars, and the grey markers correspond to field stars. The colour of each marker indicates the median membership probability. The grey vertical line represents the colour index cut for variability identification.}
    \label{statprob}
    \end{figure*}

   If a perturbation led to a successful identification of cluster member stars, the membership probability of each star was logged. After 5000 iterations, we computed, for each star, the mean, median, and standard deviation of the membership probabilities (Fig.~\ref{statprob}, left two panels). We identified 180 members with a median membership probability exceeding 0.8. Cross-matching with the $p>0$ and $m_\mathrm{G}<16$ sample of \cite{2024hunt} shows that 179 of our 180 members are in common. Nineteen stars appear only in \cite{2024hunt}, and one star appears only in our work.
   
\section{Isochrone-cloud fitting of NGC\,6866}
\label{sec:icloud}

    \subsection{Stellar evolution models}

    After obtaining the observed CMD through membership identification, our next step is to construct theoretical CMDs with well-defined input physics and their distributions. To achieve this, we employed two sets of isochrone models: MIST 1.2 \citep{dotter2016, choi2016}, which is based on the \texttt{MESA} stellar evolution code \citep{paxton2011,2013paxton, 2015paxton, 2018paxton, 2019paxton}, and the PARSEC 2.0 isochrones, based on the Padova and Trieste Stellar Evolution Code \citep{1993bressan,2000girardi, bressan2012, 2022nguyen}. 
     We chose to compare these two sets of isochrone models specifically for NGC\,6866 because, despite both providing rotating models for intermediate-mass stars, they treat several critical aspects of input physics differently.
     
     One major difference in the input physics is the treatment of convective overshoot mixing. The MIST isochrones use a diffusive overshoot scheme at the convective boundary, assuming that the convective mixing decays exponentially beyond it while the temperature gradient in the overshoot zone is the radiative one (see Sect.\,3.6.2 in \citealt{choi2016}). In contrast, the PARSEC isochrones adopt a step overshoot scheme, where overshoot is modelled as a parametrised extension of the convective boundary over a distance given by the overshoot parameter and with the adiabatic temperature gradient in the zone (see Sect.\,2.2 in \citealt{bressan2012}). Comparing these treatments provides valuable insights into how variations in convective core properties influence age measurements.  In stars with convective cores, such as the g-mode main-sequence pulsators in NGC\,6866, core overshoot leads to an increase of the hydrogen fuel reservoir in the core and hence to an extension of the star's main-sequence lifetime \citep[e.g.][]{1975maeder, 1978Roxburgh, 2013sse..book.....K}. The mass of the convective core serves as an extra dimension of calibration to be inferred from g-mode asteroseismology, in addition to the age estimate.

     Another important distinction in the input physics is the implementation of rotation and rotationally induced effects like angular momentum transport and element mixing. Both rely on the shellular approximation and cast transport processes into a diffusive formalism, but with different implementations. In the MIST models, MESA computes diffusion coefficients for a set of rotationally induced instabilities including dynamical and secular shear, Solberg–Høiland instability, Eddington–Sweet meridional circulation approached as a diffusive process, and Goldreich–Schubert–Fricke instabilities \citep[e.g.][]{2013paxton}. All these effects are assumed to lead to particular mixing levels. Their diffusion coefficients are added to those due to convection, semiconvection, thermohaline mixing and microscopic diffusion; this “grand sum” enters both the angular-momentum and composition diffusion equations, which are computed and solved at each timestep along the evolutionary tracks \citep{choi2016}. 
     PARSEC v2.0 also adopts a scheme in which the total diffusion coefficient is written as the sum of contributions from turbulent/convective mixing, shear instability (using the framework by \citealt{talon&zahn1997}) and meridional circulation, the latter treated diffusively following \citet{chaboyer1992} rather than as an explicit advective flow \citep{2022nguyen}. PARSEC v2.0 solves a single diffusion equation that couples nuclear burning with turbulent and rotational diffusion. For the intermediate-mass stars in NGC\,6866, both isochrone sets therefore implement rotation-driven transport in a time-dependent diffusive framework, but with different instability prescriptions and efficiency parameters, which can lead to non-negligible differences in the internal rotation profiles and composition gradients at fixed mass, age, and surface rotation rate.

    \subsection{Construction of an isochrone cloud}
    Both isochrone grids account for stellar rotation and adopt synthetic photometry with the \emph{Gaia} DR3 passbands. Based on the \emph{Gaia} extinction law, the extinction in the \emph{Gaia} photometric passbands ($A_\mathrm{m}$, with $m$ representing the bands $\mathrm{G,G_{BP},G_{RP}}$) for targets situated in the CMD between  $-0.06 < (\mathrm{G_{BP}-G_{RP}})_0 < 2.5$ can be calculated using $k_\mathrm{m} = A_\mathrm{m}/A_0$, where $A_0$ represents the extinction at $\lambda = 550\,\mathrm{nm}$ \citep{danielski2018}. The extinction coefficient $k_\mathrm{m}$ is given by a polynomial of $X$,
    where $X$ is the magnitude in the desired band and the coefficients are provided by \cite{riello2021} and \cite{fitzpatrick2019ApJ}.\footnote{\url{https://www.cosmos.esa.int/web/gaia/edr3-extinction-law\#}}

    To construct a theoretical isochrone, we need three parameters: age, extinction at 550\,nm ($A_0$), and initial critical rotation value.
    After adding the extinction and distance effects, we obtain an isochrone in the observational regime. We created a synthetic cluster CMD for each initial $v/v_\mathrm{crit}$ value by sampling the isochrone using the power-law initial mass function (IMF) taken from \citep{1955salpeter,kroupa2001} and adding a random scatter with standard deviation equal to the one deduced from the scatter in the observed CMD.
    
    In reality, stars in an open cluster do not have the same initial rotation rates, partially resulting in the extension of the MSTO. To address this variance, we generated a total amount of 72 different kinds of distribution of $v/v_\mathrm{crit}$, covering the range from $v/v_\mathrm{crit}=0.0$ to 0.9, using the criteria described in the Appendix~\ref{appen:fractions}. For each of the 72 possible distributions, we generated a population of 1000 stars, a number selected intentionally larger than the number of cluster members to avoid small-number statistics. The population is split according to the fraction of each rotation rate. For each subset, we sampled a corresponding number of stars from the IMF. By combining all the sampled subsets, we construct a synthetic cluster CMD with a specified age, extinction ($A_0$), and a distribution of initial rotation speeds. This synthetic CMD, which includes stars rotating at different initial speeds, is referred to as an “isochrone cloud”, a concept introduced by \citet{Johnston2019} in the context of binary modelling.

    It is worth clarifying that the initial rotation in both the MIST and PARSEC models is a parameter introduced to initiate the evolution as solid-body rotation at the ZAMS (Sect. 3.5 in \citealt{choi2016} and Sect. 2.4 in \citealt{2022nguyen}). Our isochrone–cloud inference constrains the distribution of this ZAMS parameter by optimally reproducing the present–day CMD morphology under the adopted input physics. Hence, this parameter should not be over-interpreted as the true spin distribution of actual stars at birth, but rather as an effective ZAMS distribution resulting from our chosen framework.

    \subsection{Comparison with observation}

    \begin{figure*}
    \centering
        \includegraphics[width=\linewidth]{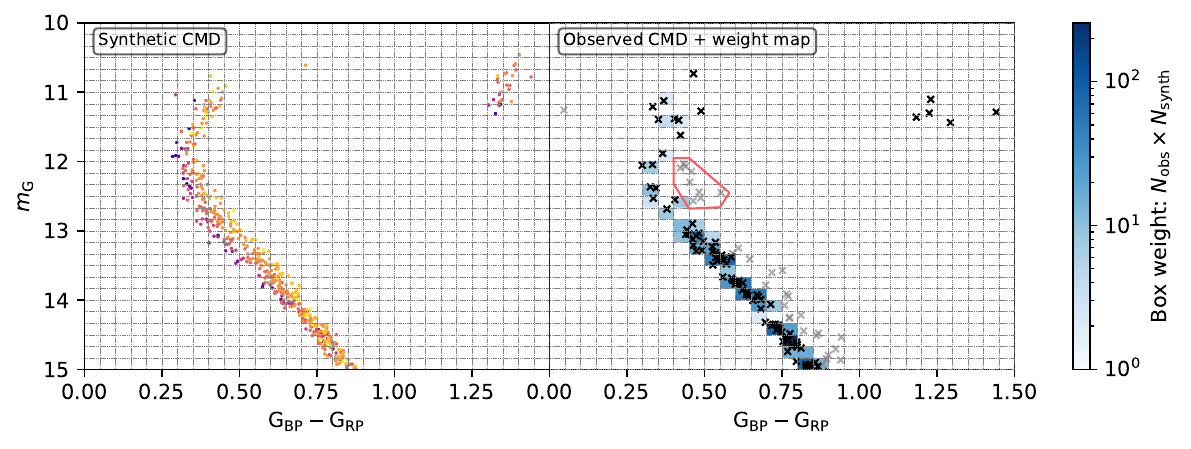}
    \caption{Colour-magnitude diagrams illustrating the comparison process using a binned multiplicative sum. The grey dashed lines in both panels represent the binning grid used in our fitting. The left panel shows a synthetic CMD using the MIST isochrone grid. The input parameters are derived from the best-fitted MIST isochrone cloud. The colour code represents the initial critical rotation of the star. The markers in the right panel show the observational CMD. The grey markers represent photometric binaries and a blue straggler star, removed before the fitting. The red polygon highlights the removed stars near the eMSTO. The colour code on the right panel represents the weight of each box, equal to the product of the number of observed stars $N_\mathrm{obs}$ and the number of synthetic stars $N_\mathrm{synth}$.}
    \label{binning}
    \end{figure*}

    The stellar models assume single stars. We thus removed photometric binaries and a blue straggler star from the observational CMD. The criterion adopted to identify photometric binaries is informed by empirical experience and guided by the CMD morphology. However, the choice was cautiously made near the MSTO due to its critical role in age determination. In the cluster CMD, a group of stars near $\mathrm{G_{BP}-G_{RP}}\approx 0.5$ and $m_\mathrm{G}\approx 12.2$ were removed, thereby yielding a reasonable eMSTO morphology, as highlighted in the right panel of Fig~\ref{binning}. For the lower-main-sequence stars, the width of the CMD after removal is narrow, and possibly unresolved binarity within the scatter is properly handled by our comparison method, limiting the impact of potential unresolved binaries on the age determination.
    
    Next, we compared the synthetic CMD with the observed CMD of NGC\,6866. As shown in Fig.~\ref{binning}, this comparison was done by binning both CMDs using the same $30\times30$ rectangular grid. This value was selected because the box size in $\mathrm{G}_{\mathrm{BP}}-\mathrm{G}_{\mathrm{RP}}$ is 0.05, slightly larger than the calculated scattering of the observed CMD. We needed the boxes to be small enough to distinguish between two similar isochrone clouds, while being large enough to avoid undersampling and prolonged computation. The selection and the robustness of this bin size is discussed in Appendix~\ref{appen:binsize}. For each rectangular box, we multiplied the number of observed stars by the number of synthetic stars falling into that box. We then summed the products of all the boxes. This multiplicative sum,
    \begin{equation}
        S = \sum_i N_{i,\mathrm{obs}}\times N_{i,\mathrm{syn}},
    \end{equation}
     with $i$ meaning the $i^\mathrm{th}$ box in both CMDs, serves as the score for the given set of input parameters, including age, $A_0$, and the initial rotation distribution. An example of this comparison method is shown in the two panels of Fig.~\ref{binning}. 

    We generated synthetic CMDs ranging from $\log_{10}(\mathrm{Age/yr}) = 7.6$ to $9.5$ with a step of 0.02\,dex, and $A_0 = 0.2\,\mathrm{mag}$ to $ 0.8\,\mathrm{mag}$ with a step of 0.01\,mag. These parameter ranges cover the entire age and extinction spread reported in the literature \citep{2022cantat-gaudin, garcia2014, brogaard2023, Kharchenko2005, bossini2019, 6811age1, bostanc2015}. To find the best-fitting age, $A_0$, and critical rotation distribution for NGC\,6866, we calculated the scores for all 410,400 combinations. Since generating the synthetic CMD involved random noise, we repeated the fitting process 400 times, each time adding a Gaussian perturbation to the observed CMD based on the same standard deviation added to the theoretical ones. The 400 sets of best-fit parameters from this process were recorded to assess the optimal fit and the statistical uncertainties.

    \subsection{Isochrone cloud fitting result}

\begin{figure*}
    \centering
    \includegraphics[width=0.48\textwidth]{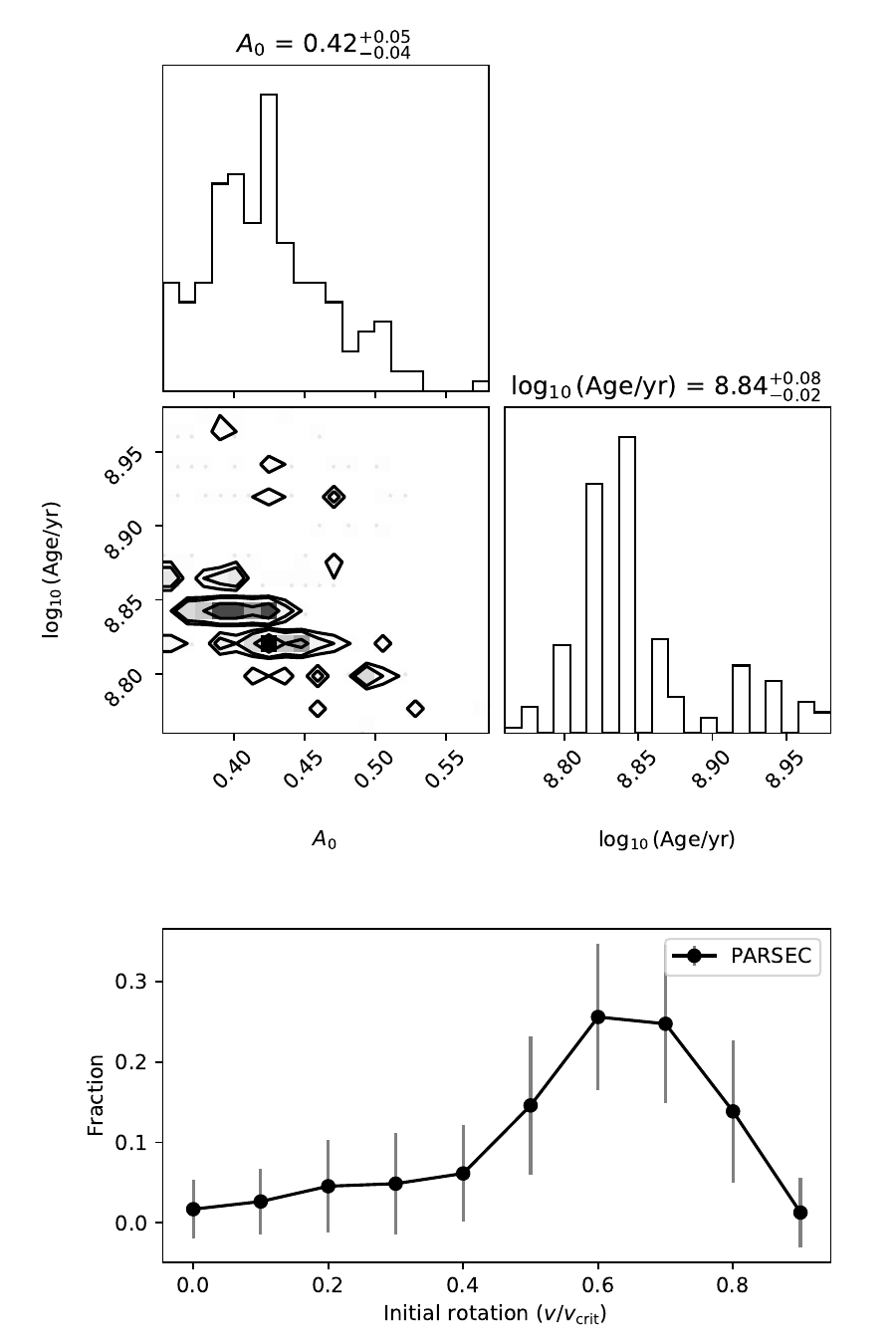}
    \includegraphics[width=0.48\textwidth]{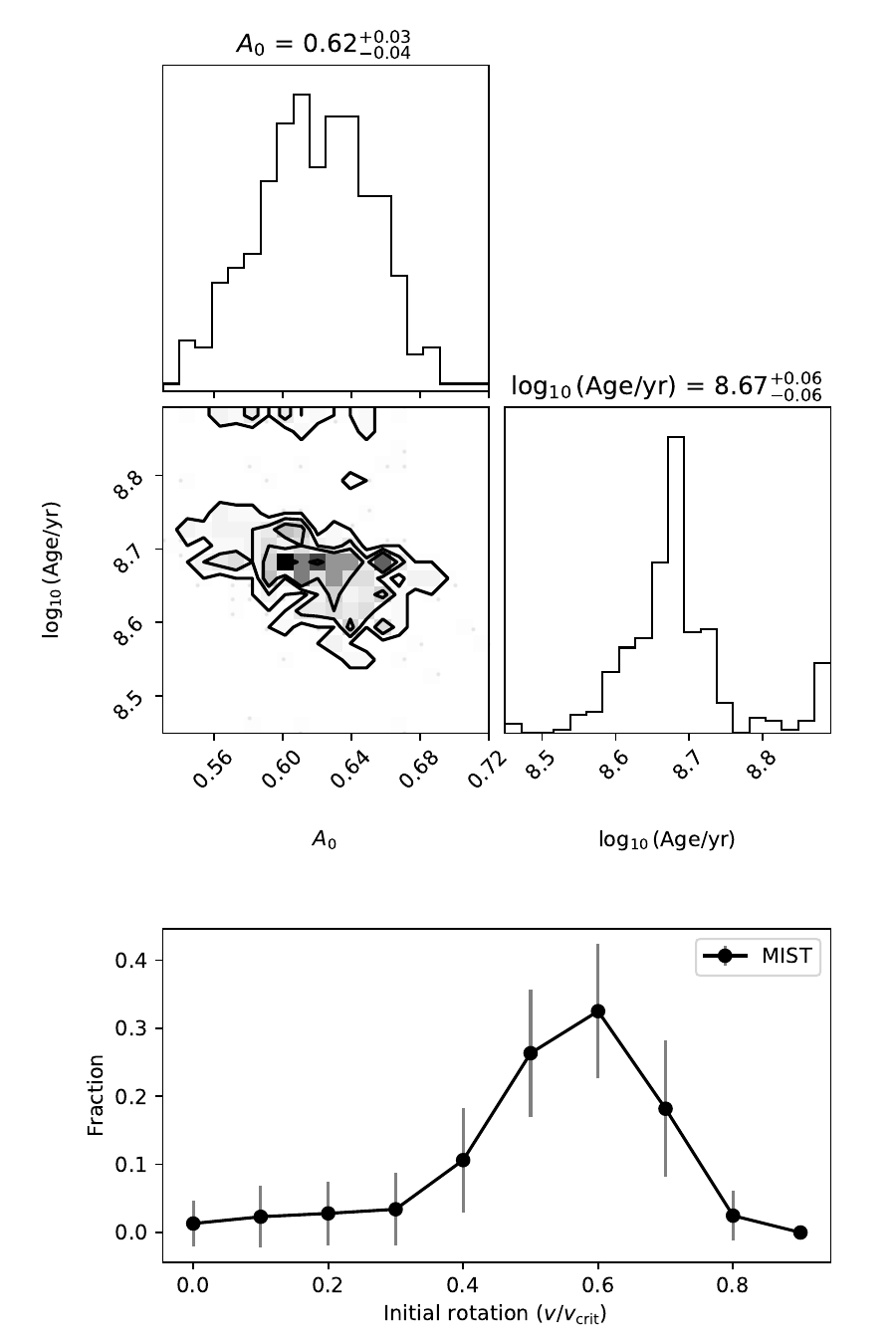}
    \caption{Best-fit isochrone cloud parameters for NGC\,6866 using the PARSEC (left) and MIST (right) isochrone models. The upper panels display the corner plots of age and $A_0$, and the lower panels show the distribution of initial critical rotation.}
    \label{fig:isochrone_combined}
    \end{figure*}

    We derived the best-fit values for the age, extinction, and the distribution of the population’s initial rotation. Additionally, the standard deviation of the best-fit parameters from the perturbation provides the statistical uncertainty of our measurements, as shown in Fig.~\ref{fig:isochrone_combined}.
    Using the MIST isochrones, the derived age of NGC\,6866 is $\log_{10}(\mathrm{Age/yr})=8.67^{+0.06}_{-0.06}$, corresponding to $467^{+70}_{-50}\,\mathrm{Myr}$ and the extinction $A_0=0.62^{+0.03}_{-0.04}$, corresponding to $E(\mathrm{B-V})=0.20^{+0.01}_{-0.01}$, taken $R_\mathrm{V}=3.1$ from \cite{ccm89} and \cite{fitzpatrick1999}. From the PARSEC isochrones, we find an older age $\log_{10}(\mathrm{Age/yr})=8.84^{+0.08}_{-0.02}$ ($690^{+140}_{-30}\,\mathrm{Myr}$) and a correspondingly lower extinction $A_0=0.42^{+0.05}_{-0.04}$ ($E(\mathrm{B-V})=0.14^{+0.02}_{-0.01}$). 
    
    \begin{figure*}
    \centering
        \includegraphics[width=\linewidth]{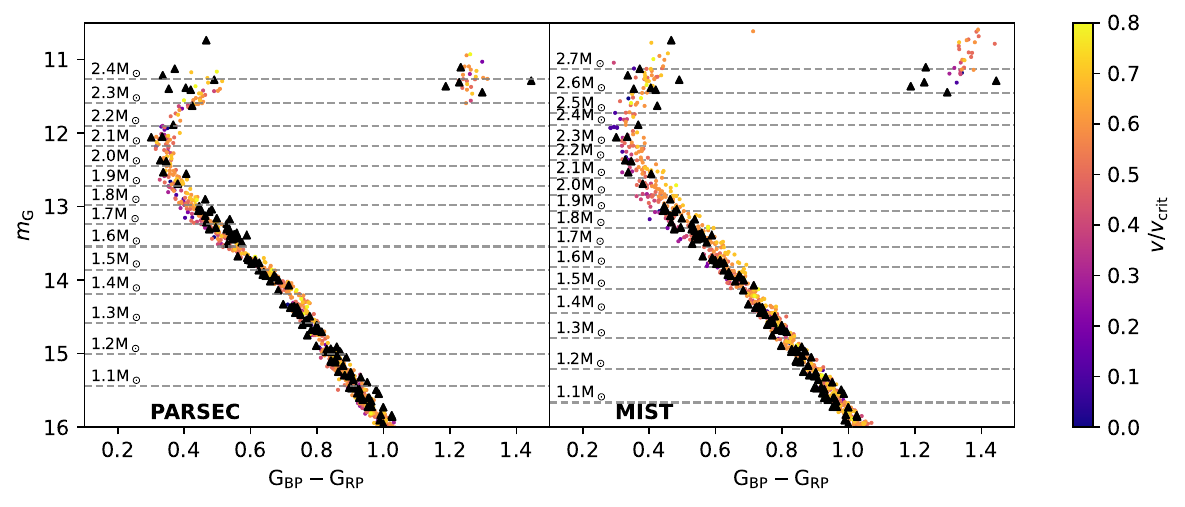}
    \caption{Synthetic CMDs based on the best-fit age, extinction ($A_0$), and initial rotation distributions. The left and right panels show the results for the PARSEC and MIST isochrone models, respectively. In each panel, the black markers represent the observed single stars in NGC\,6866, while the coloured markers represent the synthetic stars. The colour of the synthetic stars indicates their initial rotation, the same as in Fig~\ref{binning}. The grey dashed line marks the initial mass of the isochrone.}
    \label{isochrone_compare}
    \end{figure*}

    Our isochrone–cloud fitting delivers the cluster’s effective initial rotation distribution based on our adopted input physics: the upper main sequence of NGC\,6866 is dominated by fast rotators, with a peak at $v/v_\mathrm{crit}\simeq 0.6$. Rapid rotation is thus prevalent near the MSTO, consistent with rotation-based interpretations of eMSTOs and split main sequences (e.g. \citealt{DAntona2015,2019gossage}). Field A–F stars show lower-peaked distributions of their rotational velocities (e.g. \citealt{ZorecRoyer2012, Aerts2026}). In particular, population-level $\gamma$\,Dor inferences for field stars peak near $\sim\!0.25$ of the critical rotation rate (e.g. \citealt{mombarg2024b}). NGC\,6866 exhibits a distinctly faster rotating distribution for its cluster members.
    This is, to our knowledge, the first determination of an open cluster’s initial rotation distribution anchored by \emph{Gaia} photometry, providing an empirical benchmark for stellar evolution as a population. The result is most sensitive to stars in and just above the MSTO and remains model-dependent (e.g. gravity darkening, inclination, overshoot prescription); we therefore compare it directly with the seismic constraints in Sect.~\ref{sec:asteroseismology}.
    
    The two best-fit isochrone clouds are compared in Fig.~\ref{isochrone_compare}. It reveals that the MIST isochrone cloud fits the morphology of the main sequence and MSTO well, but does not match the red giant stars. In contrast, the PARSEC isochrone cloud provides a better fit for the red giant stars, but not for the MSTO region. Interestingly, the MSTO mass of $2.7\,\mathrm{M_\odot}$ from the MIST isochrone aligns with the mass measurements from previous studies \citep{brogaard2023}, where the asteroseismic masses of the red giants are approximately $2.8\,\mathrm{M_\odot}$. This disagreement between different isochrone models fitted to NGC\,6866 was also reported by the same study \citep{brogaard2023}.
 
    \section{Stellar variability in NGC\,6866}
    \label{sec:variability}

    To assemble the asteroseismic targets in NGC\,6866 (exploited in Sect.~\ref{sec:asteroseismology}), we first identified and classified the variable cluster members. This enabled a direct confrontation of the isochrone-cloud results from Sect.~\ref{sec:icloud} with independent seismic inferences (e.g.\ near-core rotation and $\Pi_0$), thereby testing the impact of input physics on the age and rotation rate derived from the isochrone-cloud fitting.

    \subsection{Target selection and frequency analysis}

    We classified stellar variability on the main sequence and on the red giant branch. Since not all members in NGC\,6866 have spectroscopic effective temperature measurements, we used the \emph{Gaia} effective temperature \texttt{gsp\_phot} $T_\mathrm{eff}$ (K) and the \emph{Gaia} colour index $\textrm{G}_{\mathrm{BP}}-\textrm{G}_{\mathrm{RP}}$ \citep{2021gaiacollab, 2023gaiacollab} to fit a colour-temperature relation. We set the colour index cut at $\textrm{G}_{\mathrm{BP}}-\textrm{G}_{\mathrm{RP}}<0.739$, which corresponds to an $T_\mathrm{eff} >6500\mathrm{K}$, approximately the red edge of the main-sequence A-F pulsator instability strip \citep{dupret2004}. We classified the variability of the main-sequence stars hotter than the cut and the manually selected red giant stars. The results of the selection are shown in the right panel of Fig.~\ref{statprob}. This yielded 85 stars in total.
   
    We cross-matched the 85 target stars selected with the \emph{Kepler} observed stars catalogue from \texttt{MAST}, \footnote{\url{https://archive.stsci.edu/kepler/catalogs.html}}, resulting in 75 stars with at least one quarter of \emph{Kepler} observations. We downloaded the \emph{Kepler} Science Operations Center (SOC) pipeline light curve of all the quarters for each star, and used the \texttt{PDCSAP\_FLUX} to compute the Lomb-Scargle \citep{1976lomb, 1982scargle} periodogram. The light curve, the periodogram, and the positions in the cluster CMD served as the basis for our manual categorisation of variability into the following types:

   \begin{itemize}
    \item g-mode pulsator: The target exhibits g-mode pulsations, with groups of peaks appearing at frequencies typically below $5\,\mathrm{d}^{-1}$. Some show period spacing patterns.
    \item p-mode pulsator: The target exhibits p-mode pulsations, with frequencies above $10\,\mathrm{d}^{-1}$.
    \item Hybrid pulsator: The target exhibits both g-mode pulsations at lower frequencies and p-mode pulsations at higher frequencies.
    \item Eclipsing binary \& rotation variable: Variabilities induced by binary eclipses or ellipsoidal variability, typically showing peaks in the lower frequency domain below $5\,\mathrm{d^{-1}}$. We manually distinguished them from the g-mode pulsators by identifying harmonic frequencies of the main frequency, caused by their non-sinusoidal variations.
    \item Solar-like oscillator: Post-main-sequence oscillators with an amplitude spectrum displaying a bell-shaped group of frequencies centred around $\nu_\mathrm{max}$ \citep{Bedding2014}.
    \item Surface modulation star: The target shows surface spot rotational modulations at low frequencies, which can be fitted using a Lorentzian profile. Some stars in this category also exhibit pulsations.
   \end{itemize}

 We identified 68 variable stars in total, which covered 80\% of the 85 candidate stars, and 38\% of the entire membership brighter than $m_\mathrm{G}=16$. The catalogue of identified variable stars within our region of interest is made available at the CDS; we summarise its columns in Table~\ref{tab:columns}. Within the context of this work, we focus on the main-sequence pulsators, as the red giant oscillators were already studied by \cite{brogaard2023}.

\begin{table}[]
\caption{Explanation of the columns in the online catalogue of variable stars in NGC\,6866.}
\resizebox{\columnwidth}{!}{%
\begin{tabular}{@{}cc@{}}
\toprule
Column name           & Explanation                                   \\ \midrule
source\_id            & \emph{Gaia} DR3 source id                   \\
KIC                   & \emph{Kepler} Input catalogue (KIC) ID                  \\
pmra                  & RA proper motion measurement from \emph{Gaia} DR3       \\
pmdec                 & Dec proper motion measurement from \emph{Gaia} DR3      \\
Mean\_Probability     & Mean value of membership probability   \\
Median\_Probability   & Median value of membership probability \\
Std\_Dev\_Probability & Standard deviation value of $p_m$  \\
class &
  \begin{tabular}[c]{@{}c@{}} gamma\_dor: g-mode pulsator\\ delta\_scuti: p-mode pulsator\\ gdor\_dcst\_hybrid: hybrid pulsator\\ solar\_like: solar-like oscillator\\ eclipsing\_binary: orbital variable\\ surface\_modulation: surface rotational modulation star\end{tabular} \\ \bottomrule
\end{tabular}%
}

\label{tab:columns}
\end{table}
    
    \subsection{g-mode pulsators}

\begin{figure*}
        \sidecaption
        \includegraphics[width=12.5cm]{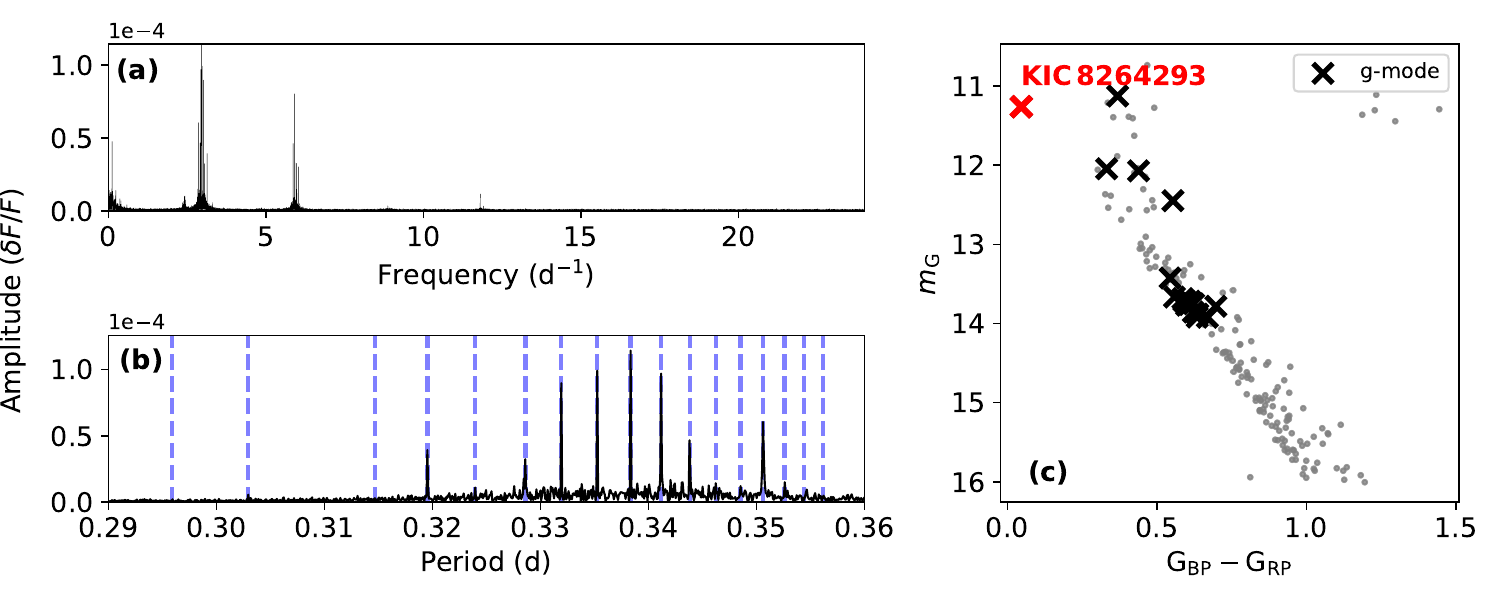}
        \caption{The pulsating blue straggler star with g-mode pulsations KIC\,8264293 in NGC\,6866. Panel (a) shows its amplitude spectrum calculated from the \emph{Kepler} light curve. Panel (b) is the zoomed-in period-amplitude diagram of its period spacing pattern. The blue dotted lines mark the fitted $l=1$ pulsation periods. Panel (c) displays the location of KIC\,8264293 in the CMD of NGC\,6866, marked by a red cross. The grey dots are the cluster members, with black crosses marking all the identified g-mode pulsators.}
    \label{gdor_example}
    \end{figure*}

    We identified 19 g-mode pulsators in NGC\,6866. The positions of these pulsators in the cluster’s CMD show a concentration within the $\gamma$\,Dor instability strip, while also extending towards the hotter part of the main sequence and reaching the MSTO (Panel c of Fig.~\ref{gdor_example}). This extension beyond the theoretical instability strip for general g-mode pulsators has been reported by \cite{2013mowlavi}, \cite{gaiacollab2023}, and \citep{mombarg2024}. Specifically for cluster g-mode pulsators, this has been reported by \cite{li2024}. These findings, along with this work, present challenges for realistic instability computations of A-F pulsators. 

    Using the \emph{Kepler} light curves, we identified clear period spacing patterns in 14 of the g-mode pulsators, including 
    the blue straggler, KIC\,8264293. This target was first reported as a blue straggler in NGC\,6866 by \cite{2009molenda-zakowicz} and confirmed by \emph{Gaia} data \citep{2021rain}. Its pulsations have been extensively studied using asteroseismic modelling \citep{2022pedersen, 2022szewczuk} treating it as an isolated single star, without taking into account its properties with respect to its host cluster. Fig.~\ref{gdor_example} shows its period spacing pattern.

\begin{figure}
        \includegraphics[width=\linewidth]{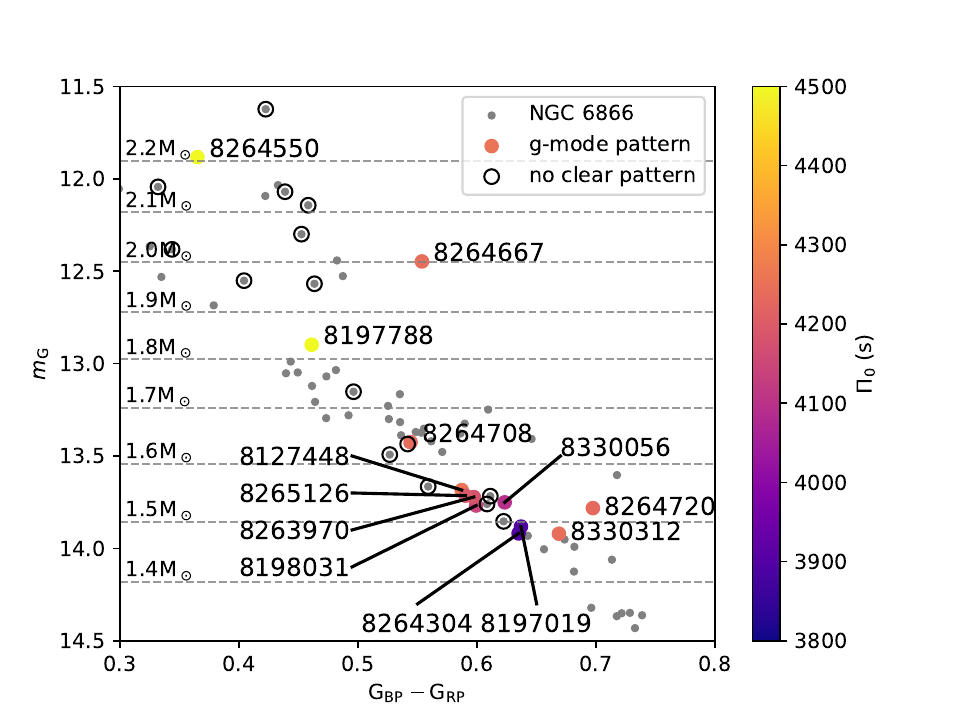}
        \caption{Zoomed-in CMD with g-mode pulsator distribution. The background grey markers are the member stars of NGC\,6866. Unfilled black circles represent the g-mode pulsators without period spacing patterns. The filled markers represent those with period spacing patterns in NGC\,6866, excluding the blue straggler KIC\,8264293. For each marker, the colour represents the measured $\Pi_0$, and the KIC numbers are labelled aside. The dashed line marks the isochronal mass of the population, based on the best-fit age, extinction and initial rotations from PARSEC.}
    \label{gdor_in_cluster} 
    \end{figure}

    For the g-mode pulsators with clear period spacing patterns, we extracted the near-core rotation frequency ($f_\mathrm{rot}$) and the asymptotic period spacing $\Pi_0$ from the catalogue by \cite{li2020_611}, who relied on the TAR to fit the patterns. Additional columns providing $f_\mathrm{rot}$ and $\Pi_0$, along with their uncertainties have been included in our online catalogue for the $\gamma$\,Dor stars of NGC\,6866. 
    Fig.~\ref{gdor_in_cluster} shows a zoomed-in CMD with the main sequence g-mode pulsators indicated and colour-coded according to their $\Pi_0$. 

    \subsection{p-mode pulsators}

    \begin{figure*}
        \sidecaption
        \includegraphics[width=12cm]{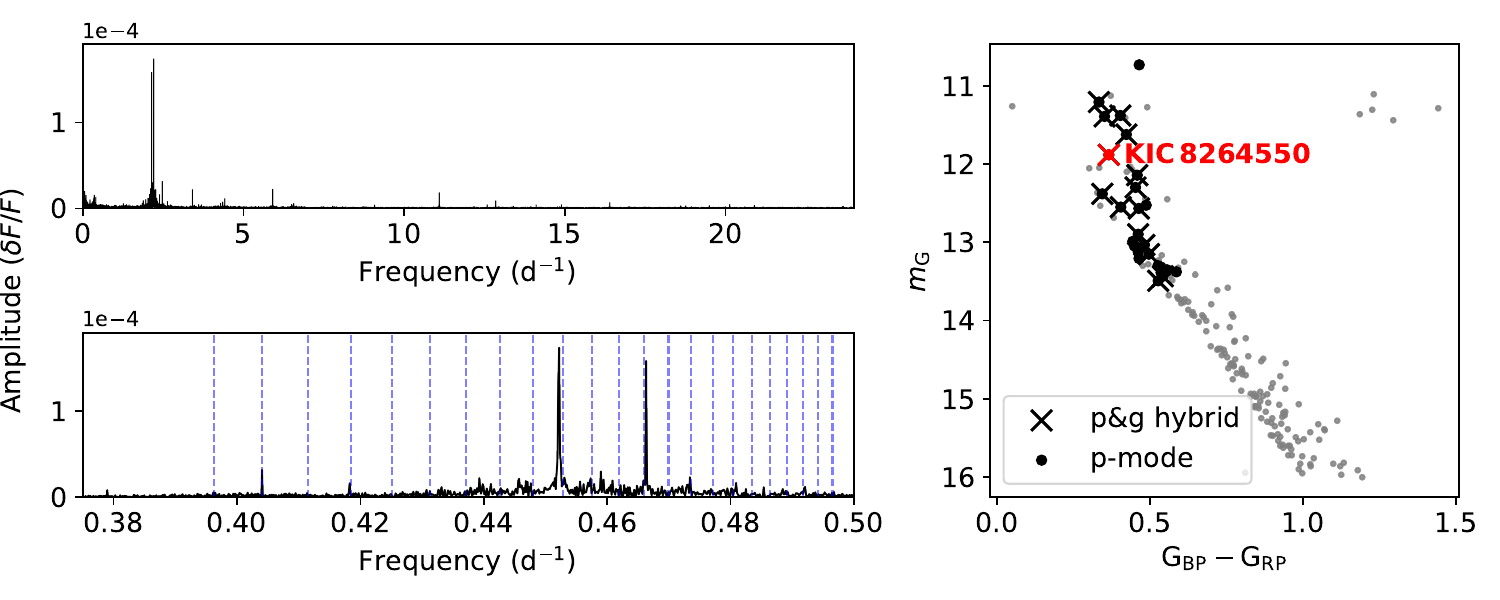}
        \caption{Hybrid pulsator KIC\,8264550 in NGC\,6866. Panel (a) shows the amplitude spectrum calculated from the \emph{Kepler} light curve where both p and g-mode can be seen. Panel (b) is the zoomed-in period-amplitude diagram of its period spacing pattern region. Panel (c) displays the location of KIC\,8264550 in the CMD of NGC\,6866, marked in red. The black markers display all the identified p-mode pulsators, and markers with an overlapping cross indicate the hybrid pulsators.}
    \label{dsct-example} 
    \end{figure*}

    We identified 27 p-mode pulsators in NGC\,6866, 14 of which also exhibited g-mode pulsations (Fig.~\ref{dsct-example}). These p-mode pulsators are predominantly located along the hotter main sequence and near the MSTO, with their temperature distribution shifting towards hotter stars compared to the $\gamma$\,Dor stars.

    As shown in Fig.~\ref{gdor_in_cluster}, the $\gamma$\,Dor region appears to be concentrated, with more than half of the stars around $\mathrm{G_{BP}-G_{RP}}=0.6$ displaying g-mode pulsations. In contrast, the region populated by $\delta$\,Sct and hybrid pulsators extends until the MSTO.

    \section{Asteroseismic modelling of NGC\,6866 g-mode pulsators}
    \label{sec:asteroseismology}
  
    We established the global parameters from isochrone-cloud fitting in Sect.~\ref{sec:icloud}. Those results suffer from model–dependent degeneracies, most notably in age and rotation because CMD fitting is sensitive to choices in input physics. The $\gamma$\,Dor pulsators allowed us to assess the interior physics properties. Their period-spacing patterns encode the near-core rotation frequency and buoyancy radius, thereby probing the interior structure in a manner complementary to the CMD. We therefore turned to forward asteroseismic modelling of the cluster’s g-mode pulsators to (i) obtain asteroseismic masses and ages, (ii) test (in)consistencies with the isochrone-based results, and (iii) assess how assumptions about coevality affect the inferred internal rotation and core properties.

    \subsection{Observations and model grid}
    
    We performed asteroseismic modelling by matching the asteroseismic parameters with a pre-computed stellar model grid. Four observables were used as input: the g-mode asymptotic period spacing ($\Pi_0$) and the near-core rotation rate ($f_\mathrm{rot}$), both measured from the period-spacing patterns under the assumption of the TAR, together with the effective temperature ($T_\mathrm{eff}$) and luminosity ($L$). The effective temperatures were derived from the colour–temperature relation described in Sect.~\ref{sec:variability}. Luminosities were computed from the \emph{Gaia} DR3 parallax ($\varpi$), the \emph{Gaia} bolometric correction ($\mathrm{BC}_\mathrm{G}(T_\mathrm{eff})$; \citealt{creevey2023}), the extinction in the \emph{Gaia} $G$ band ($A_\mathrm{G}$) measured from \emph{Gaia}, and the solar reference magnitudes ($M_\mathrm{G,\odot}=4.66\,\mathrm{mag}$, $M_\mathrm{bol,\odot}=4.74\,\mathrm{mag}$; \citealt{creevey2023}). We note that after calculating the possible $\mathrm{BC}_\mathrm{G}(T_\mathrm{eff})$ across the temperature range of our $\gamma$\,Dor stars ($6900\,\mathrm{K}<T_\mathrm{eff}<7500\,\mathrm{K}$) and, consistent with Fig.~8 of \citet{andrae2018}, we found the variation to be negligible.
    
    To avoid biases in $T_\mathrm{eff}$ and $L$ from unresolved multiplicity, we restricted the seismic modelling to single-star members, excluding the blue straggler KIC\,8264293 and the photometric binary KIC\,8264667. The final modelling sample comprises twelve $\gamma$\,Dor pulsators.

    The \texttt{MESA} model grid we used to fit the observations above is the one computed by \cite{2025li_2516_2} to analyse the g-mode pulsators in NGC\,2516. The envelope mixing is held constant and computed with a uniform viscosity profile set to $10^7\,\mathrm{cm^2s^{-1}}$. This value was calibrated using $\gamma$\,Dor stars near the terminal age main sequence (TAMS), as described in \cite{Mombarg2023}. The range of parameters of this dedicated grid was set with initial critical rotation $v/v_\mathrm{crit}$ from 0.0 to 0.5 with a step of 0.1, and the stellar mass ranges from 1.5\,$\mathrm{M_\odot}$ to 5\,$\mathrm{M_\odot}$ with a step of 0.05. Overshoot mixing near the convective boundary is implemented using an exponentially decaying scheme with three values $f_\mathrm{ov}=$ 0.005, 0.015, and 0.025, as detailed in Sec.\,3.1 of \cite{2025li_2516_2}. The final model grid used for the fitting procedure contains a total of 1,877,042 grid points.

    \subsection{Fitting methods}
\begin{figure}
        \includegraphics[width=\hsize]{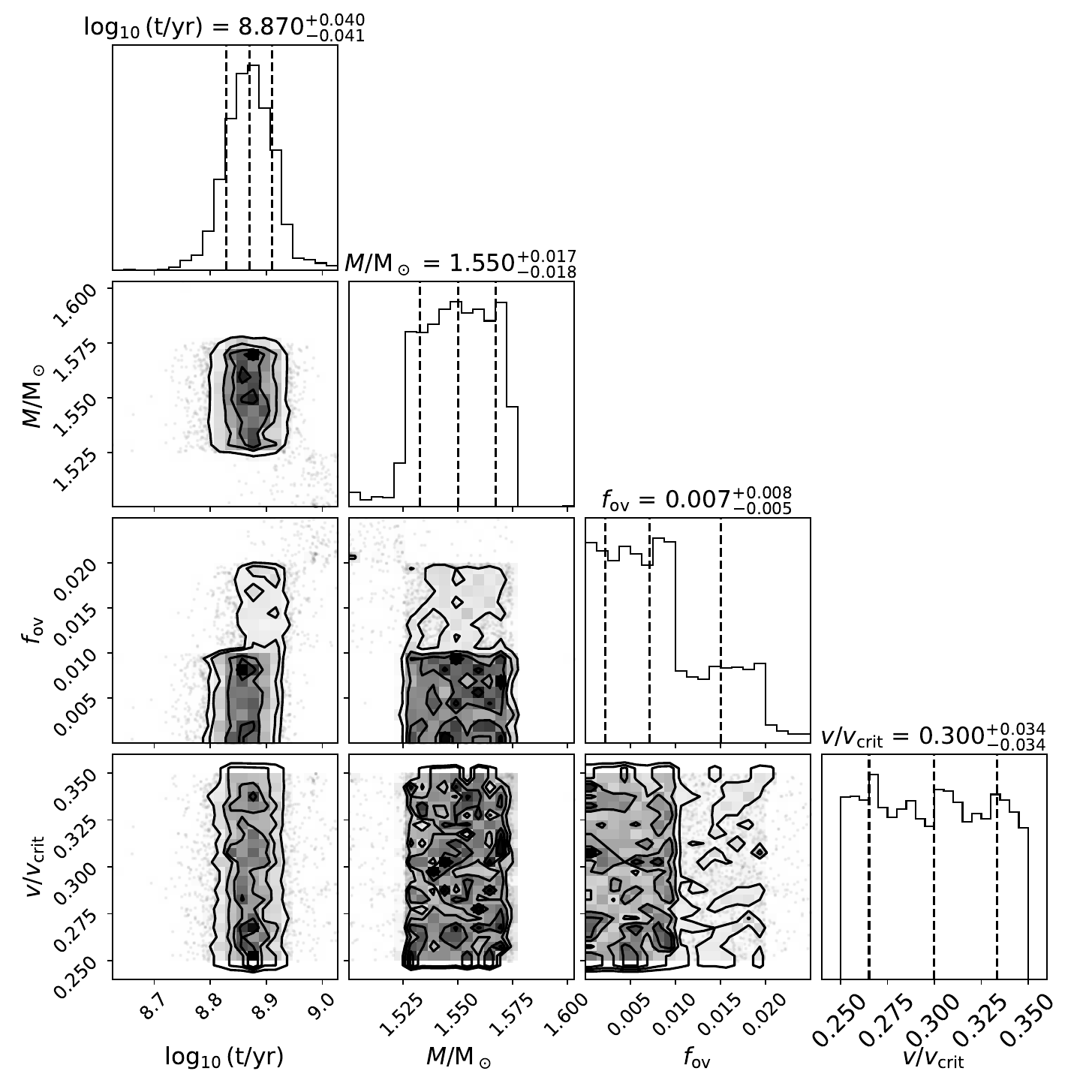}
        \caption{Result for the MCMC model fitting process for pulsator KIC\,8264708.}
    \label{model_example}
    \end{figure}

    Based on this extensive model grid, we performed grid-based asteroseismic modelling using two distinct approaches. In the first approach, we allowed each g-mode pulsator to have an independent age, just as in the case of asteroseismic grid modelling of field stars. For each of the twelve stars, we conducted Markov Chain Monte Carlo (MCMC) simulations, with the likelihood function defined as 
    \begin{equation}
    \label{eq:ll_diagonal}
    \ln \mathcal{L}(\boldsymbol{\theta}\,|\,\mathbf{y})
    = -\tfrac{1}{2}\sum_{i=1}^{k=4}\!\left[\frac{y_i-\mu_i(\boldsymbol{\theta})}{\sigma_i}\right]^2
      -\tfrac{1}{2}\sum_{i=1}^{k=4}\!\ln\!\big(2\pi\,\sigma_i^2\big)\,.
    \end{equation}
    Here $\boldsymbol{\theta}=\{M,\,f_\mathrm{ov},\,\log_{10}(\mathrm{Age}),\,(v/v_\mathrm{crit})\}$ stands for the vector of the free parameters to estimate. Via a K-D tree interpolation of the model grid, the model parameters yielded a prediction of the observables $\boldsymbol{\mu}=\{T_\mathrm{eff},\,L,\,\Pi_0,\,f_\mathrm{rot}\}$. The vector of the observations and $1\sigma$ uncertainties is
\[
\mathbf{y}=\big(T_\mathrm{eff},\,L,\,\Pi_0,\,f_\mathrm{rot}\big)_{\mathrm{obs}},
\qquad
\boldsymbol{\sigma}=\big(\sigma_{T_\mathrm{eff}},\,\sigma_L,\,\sigma_{\Pi_0},\,\sigma_{f_\mathrm{rot}}\big).
\]
    An additional important astrophysical quantity, the convective core mass ratio ($M_\mathrm{cc}/M$), was subsequently retrieved through interpolation in the model grid using the best-fit parameter vector $\boldsymbol{\theta}$. Figure~\ref{model_example} provides an example of the MCMC fitting results for KIC\,8264708. The best fit model shows an age of $\log_{10}(\mathrm{Age/yr})=8.87\pm0.04$, consistent to the age derived from PARSEC isochrone-cloud fitting
within the uncertainty. The mass $M=1.55^{+0.02}_{-0.02}$ is slightly smaller than the isochronal mass, as shown in Fig.~\ref{gdor_in_cluster}.
    
    For the second approach, we enforced a common age across all twelve pulsators, leveraging the fact that they belong to the same open cluster. The total summed log-likelihood is

    \begin{equation}
    \label{eq:ll_joint}
    \ln \mathcal{L}_\mathrm{tot}(\theta)
    =\sum_{i=1}^{N}\ln \mathcal{L}_i\!\left(\boldsymbol{\theta}_i,\,\log_{10}(\mathrm{Age}) \right),
    \end{equation}
    with $N$ representing the 12 pulsators and $\boldsymbol{\theta}_i=\{M_i,\,(f_{\mathrm{ov}})_i,\,(v/v_\mathrm{crit})_i\}$ representing the model parameters for the i-th star except the uniform age.

    \subsection{Results of seismic modelling}
    
\begin{figure}
    \includegraphics[width=\hsize]{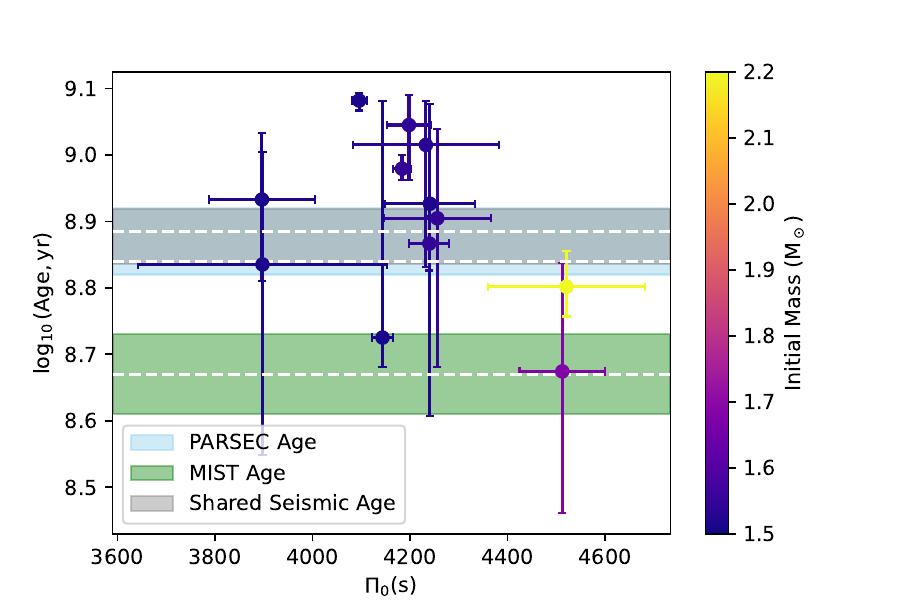}
    \caption{Age comparison from our different methods. The estimated age is plotted against the measured $\Pi_0$. The dots mark the $\Pi_0$ and asteroseismic age of the individual$\gamma\,\mathrm{Dor}$ stars, with the colour representing their fitted asteroseismic mass. The grey shade represents the asteroseismic age determined through the shared-age approach, and its uncertainty range. The sky-blue and green shades represent the uncertainty ranges of the isochronal ages using PARSEC and MIST, respectively. The upper uncertainty limit of PARSEC age and the shared seismic age overlap.}
    \label{age_compare}
    \end{figure}

    We compare the seismic ages resulting from our two approaches with the results from the isochrone-cloud fitting in Fig~\ref{age_compare}. Generally, the ages derived from the separate age fitting range from $\log t=8.7$ to $\log t=9.1$, with the two more massive g-mode pulsators beyond the classical instability strip estimated to be younger. We note that the asteroseismic ages around $\log t=9$ for the lower mass g-mode pulsators might be overestimated since their $\log L$ and $T_\mathrm{eff}$ are approaching the lower mass edge of the model grid. The shared asteroseismic age of the g-mode pulsators is $\log t=8.88^{+0.03}_{-0.05}$, overlapping with the PARSEC isochrone-cloud age.
    
    \begin{figure}
    \includegraphics[width=\hsize]{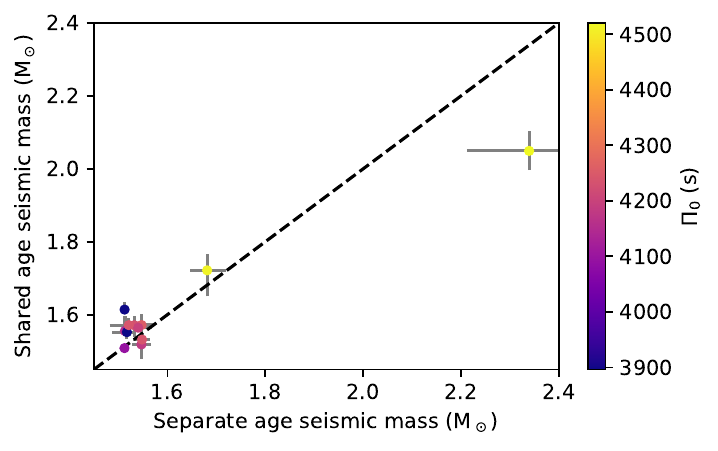}
    \caption{Mass comparison between two different fitting approaches. The colour is the measured $\Pi_0$ values.}
    \label{mass_compare}
    \end{figure}

    Figure~\ref{mass_compare} demonstrates the fitted asteroseismic mass for the g-mode pulsators using the two different approaches. These match well with the isochronal mass for the pulsators within the theoretical $\gamma$\,Dor instability strip. However, the two asteroseismic mass estimates for the hottest $\gamma$\,Dor star KIC\,8264550, having a large $\Pi_0$ and $\log L$ are discrepant. For this pulsator, we find a larger seismic mass when we do not enforce a shared age, and this value matches the isochronal mass. The other $\gamma$\,Dor star above the instability strip, KIC\,8197788 (cf.\,Fig~\ref{gdor_in_cluster}) has a similar $\Pi_0$ as KIC\,8264550, while its $\log L$ is much lower. The optimal ages fitted for these two stars differ notably, possibly affecting their fitted mass.

    As an indicator of the evolutionary stages of the pulsators, we recovered the ratio between the core hydrogen mass fraction and its initial value $X_\mathrm{c}/X_\mathrm{ini}$ from the best fit model parameters. The lower-mass pulsators are at similar evolutionary stages, with their $X_\mathrm{c}/X_\mathrm{ini}$ ranging from 0.8 to 0.6, for both the cases of using separate or shared-age fitting. The more massive KIC\,8264550 on the other hand, has its $X_\mathrm{c}/X_\mathrm{ini}\approx0.28$ and $0.40$ from separate and shared age fitting respectively, indicating that it is indeed more evolved than its siblings in line with its higher mass. For all g-mode pulsators, we also computed the specific angular momentum, $J/M$. The values for all g-mode pulsators in NGC\,6866 agree with the values for the field pulsators found by \citet{Aerts2026} and also  
    fulfil the upper limit derived in that paper for the mass regime covered here.

\begin{figure}
        \includegraphics[width=\hsize]{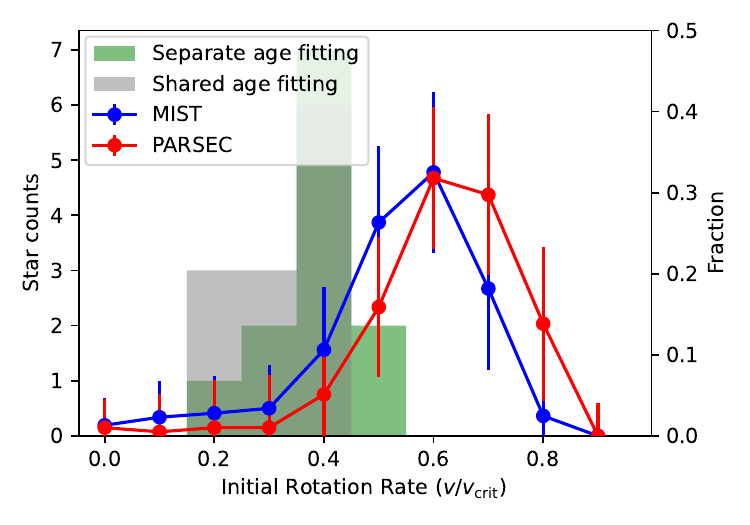}
        \caption{The results of the initial rotation rate at ZAMS $v/v_\mathrm{crit}$ from four origins. The grey and green histograms represent the count of stars with different initial $v/v_\mathrm{crit}$, measured from shared-age seismic fitting and separate age seismic fitting, respectively. The red and blue lines represent the distribution of the initial $v/v_\mathrm{crit}$, returned from the PARSEC and MIST isochrone-clouds.}
    \label{vvcrit_compare}
    \end{figure}

    In addition, we compared our effective initial $v/v_\mathrm{crit}$ values derived within the shared-age fitting to those from the isochrone cloud fitting, as shown in Fig~\ref{vvcrit_compare}. The peak of the seismic $v/v_\mathrm{crit}$ distribution determined through shared-age fitting is centred around $v/v_\mathrm{crit}=0.3$ and has a wider spread than the distribution derived from the individual age fitting, which peaks at $v/v_\mathrm{crit}=0.4$.
    Both seismic distributions occur at lower values than the isochronal distributions peaking around 0.6. We attribute this discrepancy to the difference in angular momentum transport efficiency between the two grids of isochrones on the one hand and the asteroseismic model grid on the other hand. The latter grid is tuned to represent the observed oscillation properties of both field and cluster g-mode pulsators \citep{2025li_2516_2, Aerts2026}. We do note that the 
    isochronal ages are particularly sensitive to stars on the upper main sequence near the MSTO, quite above the mass range of most g-mode pulsators in NGC\,6866. Nevertheless, the rotational slowdown in field g-mode pulsators is common to all stars covering a mass range of 1.3\,M$_\odot$ until about 2.5\,M$_\odot$
    \citep{Aerts2026}.
    
\begin{figure*}
    \centering
    \includegraphics[width=0.9\textwidth]{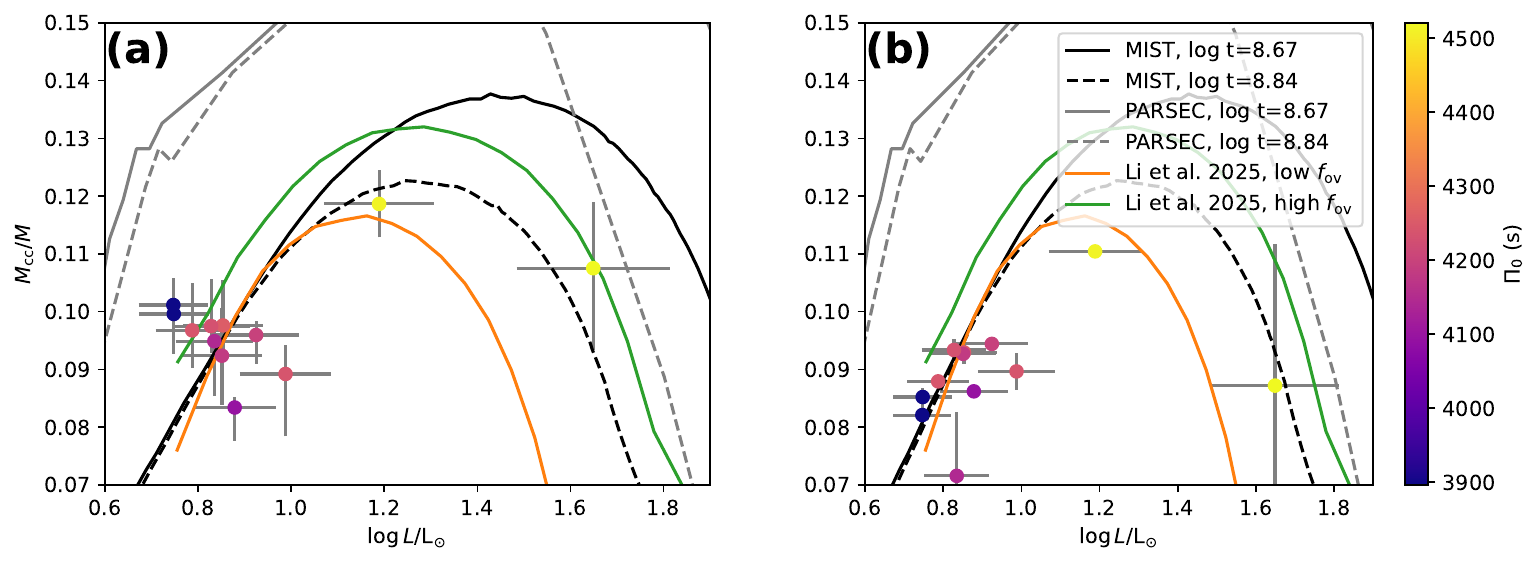}
    \caption{Convective core mass ratio $M_\mathrm{cc}/M$ recovered from grid modelling, as a function of observed luminosity for the twelve g-mode pulsators. Panel (a) shows the results of the shared-age fitting and panel (b) of the separate-age fitting. The colour of the markers represents the measured $\Pi_0$ values. The plotted lines represent the $M_\mathrm{cc}/M$ from isochrones. Grey and black lines represent the PARSEC and MIST isochrones, respectively. The dashed and solid line style represents the age of the isochrone, respectively $\log t\,\mathrm{(yr)}=8.67$ and $\log t\,\mathrm{(yr)}=8.84$ from isochrone cloud fitting. The green and orange lines represent the isochrone constructed from the model grid in \citet{2025li_2516_2}, with $\log t\,\mathrm{(yr)}=8.88$ from shared-age asteroseismic fitting. The green line uses $v/v_\mathrm{crit}=0.4$, $f_\mathrm{ov}=0.005$, while the orange line uses $v/v_\mathrm{crit}=0.4$, $f_\mathrm{ov}=0.025$.}
    \label{fig:mccm_result}
\end{figure*}

\begin{figure}
    \centering
    \includegraphics[width=\hsize]{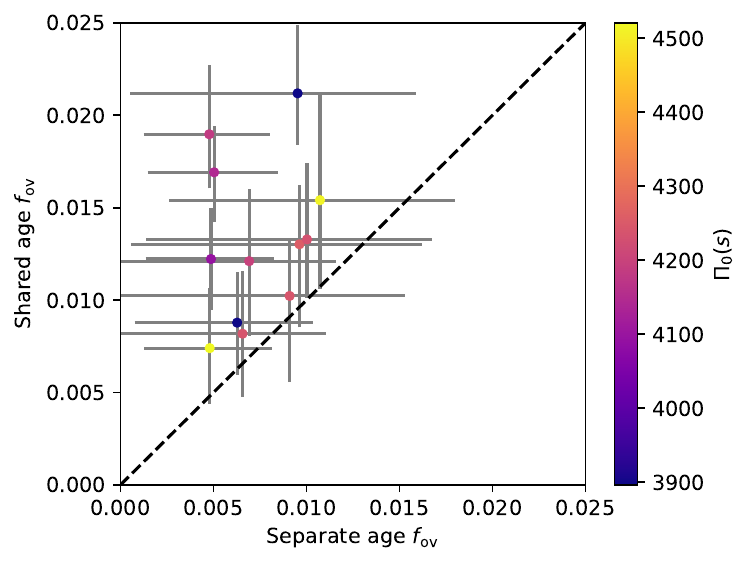}
    \caption{Comparison of the derived exponential overshoot values, $f_\mathrm{ov}$, of our two fitting approaches. The colour indicates the measured $\Pi_0$.}
    \label{fig:fov_result}
\end{figure}

    Finally, we inferred the convective core mass ratio $M_\mathrm{cc}/M$ and compared it with the values from isochrone models, as an inspection of the impact of core overshoot. Figure~\ref{fig:mccm_result} shows the $M_\mathrm{cc}/M$ as a function of observed luminosity, representing the mass sequence at a fixed age. All $M_\mathrm{cc}/M$ generally follow the trend of the MIST and asteroseismic isochrones from the model grid in \cite{2025li_2516_2}. Values derived from separate-age fitting (right panel in Fig.~\ref{fig:mccm_result}) show a more concentrated distribution, generally aligning closely with the isochrone with $f_\mathrm{ov}=0.005$. The $M_\mathrm{cc}/M$ values obtained from shared-age fitting exhibit a somewhat larger spread, which stems from a wider-spread of the core overshoot parameter $f_\mathrm{ov}$, as shown in Fig~\ref{fig:fov_result}. The uncertainties of $M_\mathrm{cc}/M$ and $f_\mathrm{ov}$ derived from shared age fitting are systematically larger than from separate age fitting.

    This discrepancy is partially methodological, since enforcing coevality removes one degree of freedom per star. In turn, star-to-star differences in $(\Pi_{0},f_{\rm rot})$ are accommodated by a broader range of core-boundary mixing, which inflates the posteriors of $f_{\rm ov}$ and hence $M_{\rm cc}/M$. On the other hand, cluster members may have star-to-star variations in near-core mixing linked to mass and rotation. Within our present grid, we therefore interpret the increased scatter of parameters in the shared-age approach as a combination of a partially methodological effect, with a non-negligible component of intrinsic physical diversity.

\section{Summary and conclusions}
\label{sec:conclusion}

    In this work, we studied the open cluster NGC\,6866 in the \emph{Kepler} field, followed by an asteroseismic analysis of its g-mode pulsating members. Methodologically, we introduced an isochrone-cloud approach that samples multiple input-physics realisations and initial-rotation distributions. We utilised the g-mode pulsator cluster members and modelled them twice, once as individual stars and again while enforcing a common age. Together, these analyses deliver a cluster-level initial rotation distribution and seismic constraints on age and near-core structure, enabling a direct confrontation between isochronal and seismic inferences. 
    
    We conducted a membership identification using a Gaussian Mixture Modelling clustering algorithm applied to the high-precision \emph{Gaia} DR3 astrometric data. We estimated the uncertainties of our membership probability using a perturbative Monte Carlo process. We found 180 high probability members with $m_\mathrm{G}<16$.

    We applied an isochrone-cloud fitting method with a simulated stellar population, enabling us to consider isochrones with different initial conditions, such as
    initial critical rotation $v/v_\mathrm{crit}$. We achieved it by simulating a synthetic cluster CMD with stars from different rotating isochrones and comparing them with observations through a binned multiplicative sum approach. We found that the cluster age and extinction from isochrone cloud fitting using PARSEC isochrones is $690^{+140}_{-30}\,\mathrm{Myr}$, with $A_0=0.42^{+0.05}_{-0.04}$. For MIST isochrones, the results were $467^{+70}_{-50}\,\mathrm{Myr}$, with a corresponding larger $A_0=0.62_{-0.04}^{+0.03}$. We also derived the distribution of the initial rotation rates for both MIST and PARSEC isochrones and found that the upper main sequence of NGC\,6866 is dominated by fast rotators, with the distributions both peaking at initial $v/v_\mathrm{crit}=0.6$ for the adopted input physics in the two used public isochrone databases. These results are robust to reasonable choices of binning and to plausible levels of unresolved, low-$q$ multiplicity; nevertheless, single-star assumptions remain an important systematic to revisit in the future.

    We used the 4-year \emph{Kepler} photometry data for variability classification and frequency analysis. Among the stars within our region of interest, we identified 19 pure g-mode pulsators and 27 p-mode pulsators, among which 14 are hybrid pulsators showing both p- and g-modes. By matching the mass at the main-sequence turn-off with asteroseismic masses for post-main-sequence stars in the literature \citep{brogaard2023}, we concluded that the PARSEC isochrones with a turn-off mass at $2.4-2.5\,\mathrm{M_\odot}$ agree better with the observations than the MIST isochrones. 
    
    The turn-off mass of NGC\,6866 occurs at the break in the specific angular momentum found for a large sample of field g-mode pulsators by \citet{Aerts2026} making the cluster an ideal laboratory for more detailed studies of transport processes.

    Among all pulsators with g~modes, 14 show clear period spacing patterns, which allowed us to determine their near-core rotation rates and asymptotic spacings $\Pi_0$.
    We obtained values in the range $3800\,\mathrm{s}<\Pi_0<4500\,\mathrm{s}$, while the {\it Gaia\/} DR3 effective temperatures cover $7000\,\mathrm{K}<T_\mathrm{eff}<8500\,\mathrm{K}$. The near-core rotation rates for the $\gamma$\,Dor stars in NGC\,6866 are between $1\,\mathrm{d^{-1}}$ and $2\,\mathrm{d^{-1}}$, which is lower than the values found for such stars in the younger open clusters NGC\,2516 \citep{li2024,2025li_2516_2} and the Pleiades \citep{Fritzewski2026}. The comparative results on internal rotation between these three open clusters are in agreement with the slowdown found for field stars by \citet{Aerts2025} and \citet{Aerts2026}. 
   
    We found relatively hotter $\gamma$\,Dor stars beyond the theoretical $\gamma$\,Dor instability strip within the eMSTO. Moreover, one hot g-mode pulsator, KIC\,8264293, is a blue-straggler in NGC\,6866, with $\Pi_0\simeq 8000$\,s and $T_\mathrm{eff}>10000\,\mathrm{K}$. We hence suggest that KIC\,8264293 is a slowly pulsating B (SPB) star having undergone a merging process, which rejuvenated the star and prolonged its main-sequence life.

    With the identified g-mode pulsators, we performed grid-based asteroseismic modelling of the $\gamma$\,Dor stars on the single-star main sequence, comparing their observed seismic diagnostics with a dedicated rotating \texttt{MESA} model grid \citep{2025li_2516_2}. We applied two fitting approaches: (i) a separate-age fit that treats each star independently, and (ii) a shared-age fit that enforces one age for all stars, leveraging their shared cluster membership. From both approaches, we derived stellar ages, masses, and initial critical rotation rates, and we compared these to the values from our isochrone-cloud fitting.

    The individual asteroseismic masses from the separate-age fits agree with the isochronal masses. However, for the hottest g-mode pulsator KIC\,8264550, the mass from the shared-age fit differs about 0.2\,$\mathrm{M_\odot}$ from the separate-age seismic mass. The shared-age fit yields a seismic age of $759^{+54}_{-82}\,\mathrm{Myr}$, consistent with the PARSEC isochrone age but not with the MIST age. \citet{brogaard2023} reported an even younger age of $t\simeq430$\,Myr from red-giant asteroseismology, about 8\% lower than our MIST isochrone-cloud age (467\,Myr) and roughly 50\% lower than our seismic age based on the g-mode pulsators. Our modelling directly constrains main-sequence core properties via $\Pi_0$ and $f_{\rm rot}$, showing that part of the tension is attributable to input-physics choices like mixing and rotational evolution, but it does not resolve potential systematics specific to red-giant inferences. Such discrepancies 
    are not unexpected; for example, \citet{2024b_fritzewski} showed that ages from red-giant asteroseismology obtained while ignoring the effects of rotation during the long main-sequence phase can differ by up to 20\% for field stars, depending on the input physics of the isochrones. These differences highlight systematic uncertainties in age-dating, pointing to the need for better calibrated stellar models across the main-sequence and red-giant phases.

    The initial critical rotation values derived from the $\gamma$\,Dor stars through asteroseismic fitting are systematically lower than the distribution obtained from isochrone cloud fitting. Possible explanations for this discrepancy are the different mass range probed and/or differences in the treatment of angular momentum transport. We also observed small differences in the internal rotation distributions derived from the two asteroseismic fitting methods. Treating stars individually resulted in a distribution peaking around $v/v_\mathrm{crit} = 0.3$, while the common-age approach produced a somewhat broader distribution. Our findings show that enforcing a common age for asteroseismic modelling in open clusters is both feasible and justified, although some stars may be merger or accretion products which cannot be explained by single-star evolutionary models.

    Similar conclusions emerged from the convective core mass measurements. While the individual-age fitting approach yielded convective core mass ratios and overshoot parameters closely aligned with those deduced from the MIST isochrone, the shared-age fitting resulted in a somewhat broader spread. Convective core mass ratios were estimated to be higher than those from the separate fitting approach, due to the higher overshoot parameter estimations. The broader $f_{\rm ov}$ and $M_{\rm cc}/M$ posteriors reflect a combination of methodology and possible intrinsic diversity
    in internal mixing leading to star-to-star variations among the cluster members. These results reinforce the value of assuming a shared age when analysing open cluster stars, as it helps mitigate potential biases arising from imposing a single value of core overshoot for an entire population as a too restrictive modelling approach
    \citep{2019johnston,Johnston2021}.

    Overall, our study demonstrates that NGC\,6866 is an optimal test bed for more detailed asteroseismic analyses. We partially addressed the challenge of adopting different initial conditions by sampling isochrone clouds with a distribution to reproduce the observed eMSTO. Similar studies of g-mode pulsations in open clusters have previously been conducted only for three younger TESS clusters, UBC-1 \citep{fritzewski2024}, NGC\,2516 \citep{li2024} and the Pleiades \citep{Fritzewski2026}. Compared to TESS, the high-precision \emph{Kepler} photometry used in this work allows more accurate measurements of stellar rotation rates and asymptotic period spacings. Furthermore, systematic discrepancies between asteroseismic and isochronal analyses -- in age, mass, and initial rotation profiles -- remain to be addressed in more detail by incorporating all identified pulsation modes in all cluster pulsators, including the SPB blue straggler.

\section*{Data availability}

Table~\ref{tab:columns}, and the g-mode asteroseismic modelling results using both approaches, are only available in electronic form at the CDS via anonymous ftp to \url{cdsarc.u-strasbg.fr} (130.79.128.5) or via \url{http://cdsweb.u-strasbg.fr/cgi-bin/qcat?J/A+A/}.

\begin{acknowledgements}

The research leading to these results has received financial support from the Flemish Government under the long-term structural Methusalem funding program by means of the project SOUL: Stellar evolution in full glory, grant METH/24/012 at KU Leuven, Belgium, as well as from the European Research Council (ERC) under the Horizon Europe programme (Synergy Grant agreement No.101071505: 4D-STAR). While partially funded by the European Union, views and opinions expressed are, however, those of the authors only and do not necessarily reflect those of the European Union or the European Research Council. Neither the European Union nor the granting authority can be held responsible for them. 
G.L. acknowledges the support of the Australian Research Council through the DECRA project DE250100773. 
This research has made use of NASA's Astrophysics Data System Bibliographic Services and of the SIMBAD database, operated at CDS, Strasbourg, France. The computational resources and services used in this work were provided by the VSC (Flemish Supercomputer Centre), funded by the Research Foundation Flanders (FWO) and the Flemish Government.

\end{acknowledgements}

\bibliographystyle{aa}
\bibliography{ref.bib}
\begin{appendix}
\section{Membership of NGC\,6866 compared to \cite{2024hunt}}
We compared our results of membership identification with previous work based on the same \emph{Gaia} DR3 astrometry data \citep{2024hunt}. For members with $m_\mathrm{G}<16$ and $p>0$ in \cite{2024hunt}, we reached 179 common member stars, with one star unique to our result and 14 stars unique to \cite{2024hunt}.

    \begin{figure}
        \includegraphics[width=\linewidth]{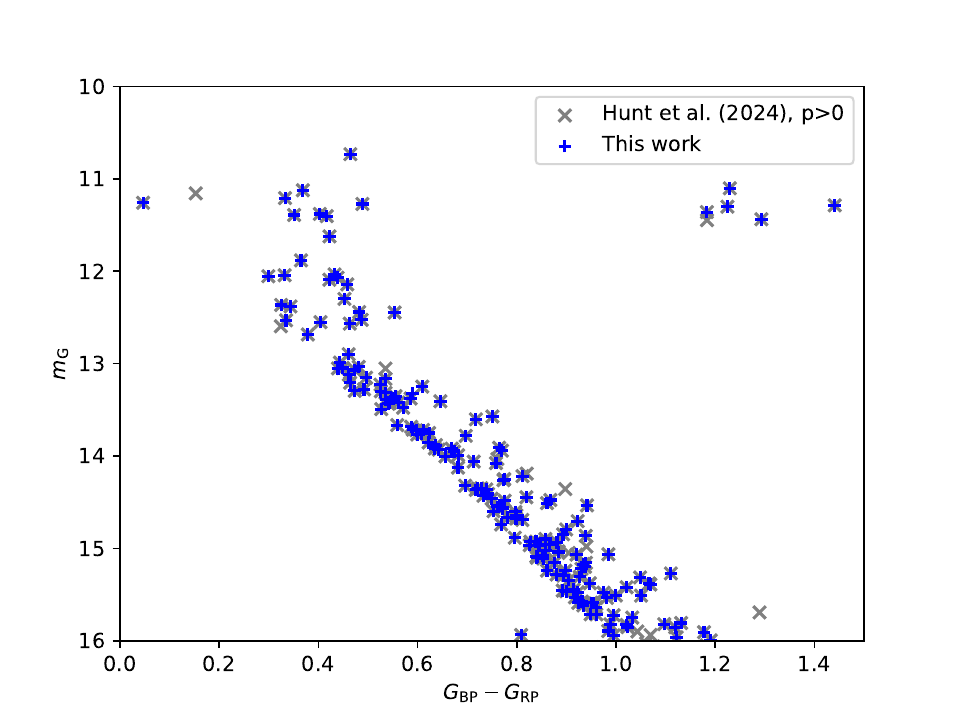}
        \caption{CMD of NGC\,6866 from this work and from the $p>0$ samples of \cite{2024hunt}.}
    \label{solarlike-example}
    \end{figure}

\section{Generation of the $v/v_\mathrm{crit}$ distribution}
\label{appen:fractions}

    We consider all the distributions which follow these five conditions:
   
	1.	Critical sampling constraint: The rotation rates, expressed as the ratio of Keplerian critical speed ($v/v_\mathrm{crit}$), range from 0 to 0.9 with a step size of 0.1.
    
	2.	Normalisation condition: The sum of the fractions of stars with different initial rotations must equal 1.
    
	3.	Multiplicity requirement: Each rotation distribution must contain at least three distinct initial rotation values.
    
	4.	Continuity condition: The distribution of initial rotation values is continuous, meaning that there cannot be a zero fraction between two non-zero fractions.

   This results in a total of 72 possible distributions.

\section{Choice of the number of bins}
\label{appen:binsize}
Demonstrated here is the robustness of different bin sizes chosen for our comparison method. We aimed to test the ability to distinguish different isochrone clouds for three different bin sizes. For a range of combinations of $A_0$ and $\log{t}$, we calculated the multiplicative sum scores $S$ with respect to three different choices of bins, including $15\times15$, $30\times30$ and $45\times45$ (corresponding to bin sizes $0.1\times0.33$, $0.05\times0.167$, $0.03\times0.11$ in $m_\mathrm{G}$ and $G_\mathrm{BP}-G_\mathrm{RP}$ magnitudes), using the PARSEC isochrone model. The distribution of the initial rotations was fixed to the best-fit PARSEC distribution for computational reasons.

    \begin{figure*}
        \includegraphics[width=\linewidth]{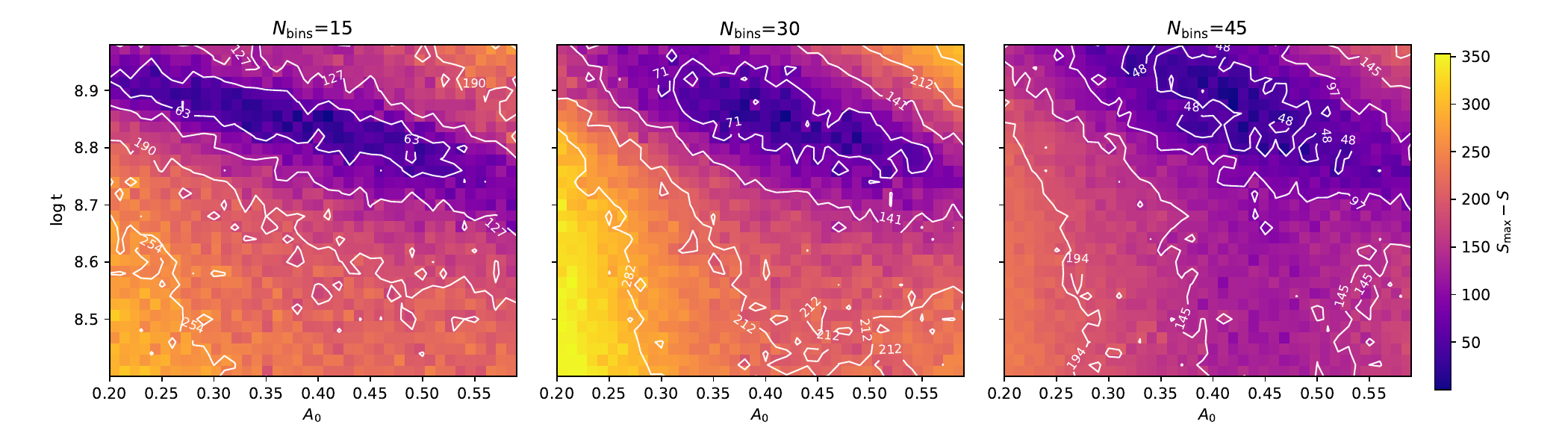}
        \caption{From left to right, multiplicative sum score $S$ calculated for different number of bins ($15\times15$, $30\times30$ and $45\times45$). Within each panel, the x and y axes represent the age and extinction parameters of isochrone clouds within the grid. The colour of each pixel represents the difference between its $S$ score and the corresponding score of the best-fit isochrone cloud $S_\mathrm{max}$. The overplotted contours trace equally spaced levels of $S_\mathrm{max}-S$ between the minimum and maximum values within each panel.}
    \label{fig:nbins}
    \end{figure*}

The results for different bin sizes can be seen in the comparisons in Fig.~\ref{fig:nbins}. A $15\times15$ grid is not able to effectively differentiate between two isochrone clouds with different $A_0$ because the bin sizes were too large. The contours of the $S$ scores are similar across a wide span of $A_0$. In practice, this would result in larger and more unreliable systematic uncertainties, as the best-fit parameter sets were less constrained. Comparing the middle and the right panels shows a roughly similar ability in distinguishing isochrone clouds, as the central contours show similar sizes. However, the $30\times30$ grid gave smaller scatter at younger ages. The computation time for the $45\times45$ grid was approximately 2.5 times that of the $30\times30$ grid. Considering both the statistical effectiveness and the calculation efficiency, we chose $30\times30$ as our grid for comparisons.

\section{Other types of variable stars in NGC\,6866}
    \subsection{Solar-like oscillators}

    We identified three solar-like oscillators in NGC\,6866, located just beyond the Hertzsprung gap. Among them, only one star, KIC\,8461659, exhibits resolved mixed $l=1,3$ modes. The other two stars show a $\Delta_\nu$ profile with $l=0,2$ modes. The observational data used to identify their variability and the distribution are shown in Fig.~\ref{solarlike-example}. All three of these solar-like oscillators were already reported and analysed by \cite{brogaard2023}.

    \begin{figure*}
        \includegraphics[width=\linewidth]{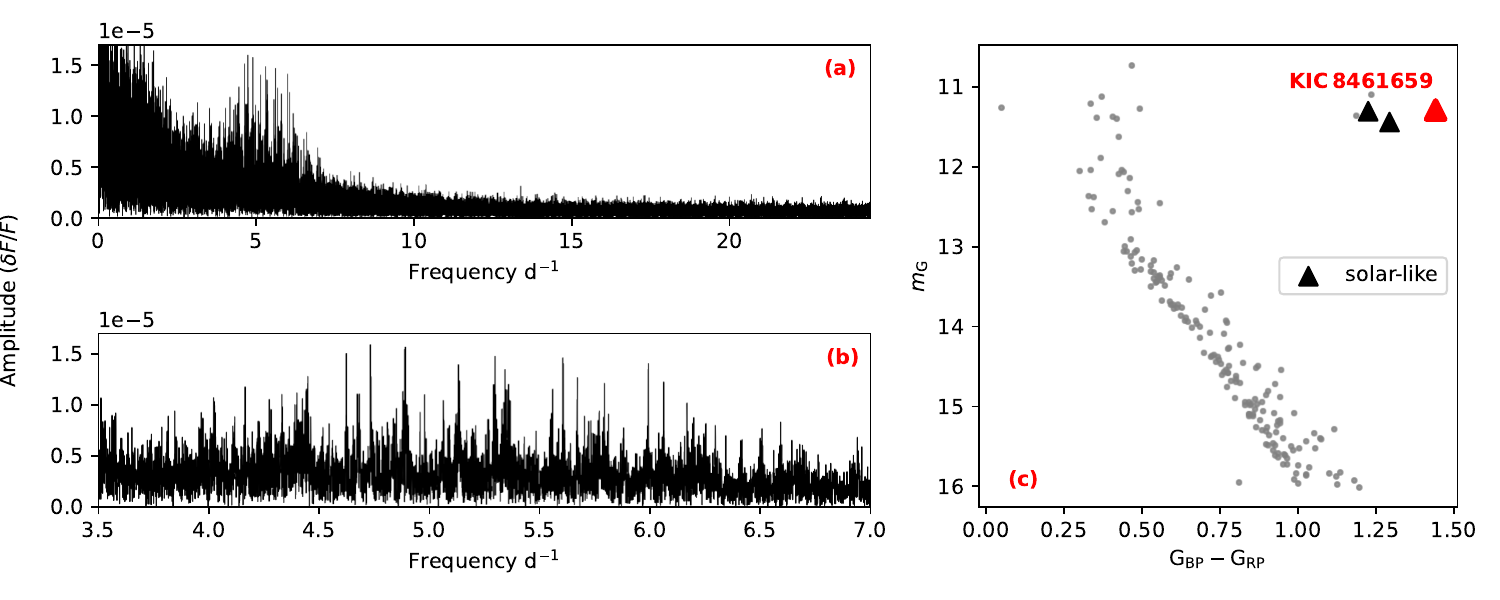}
        \caption{Solar-like oscillator KIC\,8461659 in NGC\,6866. Panel (a) shows the power density spectrum of the star, which was calculated from the light curve data measured from the \emph{Kepler} pipeline light curve. Panel (b) shows the zoomed-in power density spectrum of the $\nu_{max}$ profile containing clear mixed-mode frequencies. Panel (c) displays the location of KIC\,8461659 on the CMD of NGC\.6866, marked with a red triangle. The other two solar-like oscillators are marked with black triangles.}
    \label{solarlike-example}
    \end{figure*}

    \subsection{Orbital variables}

    \begin{figure*}
        \includegraphics[width=\hsize]{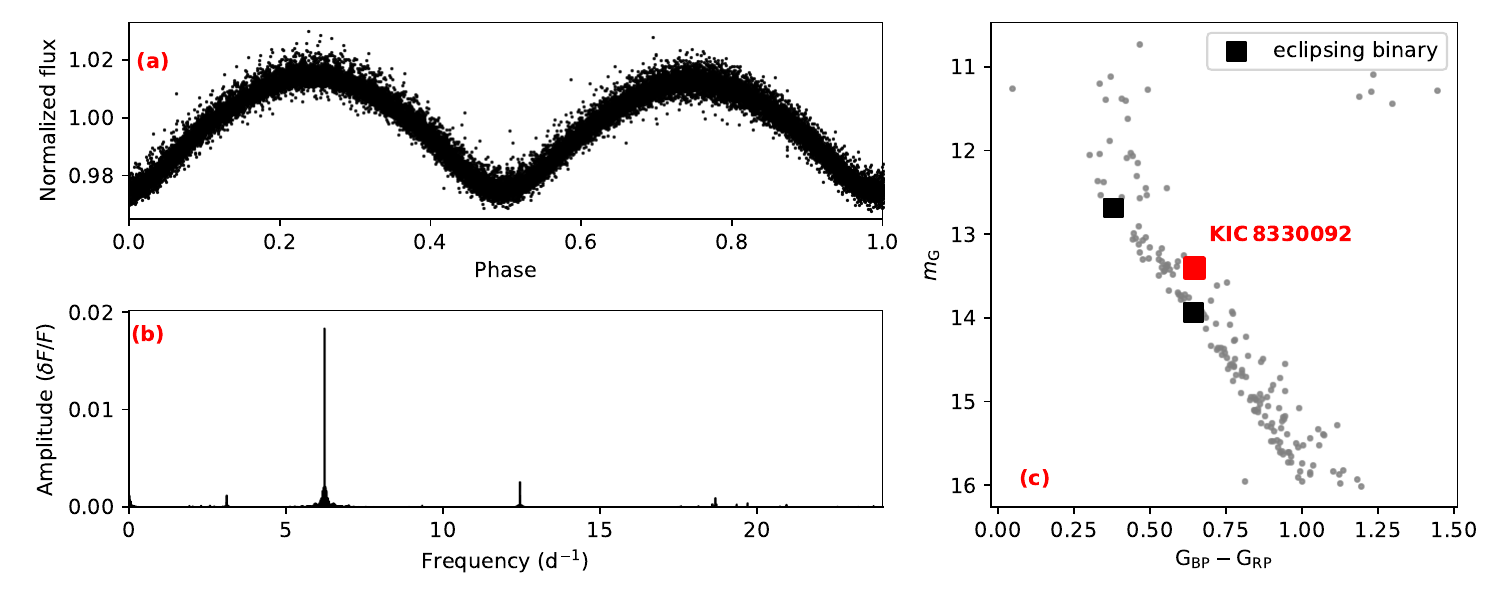}
        \caption{Eclipsing binary KIC\,8330092 in NGC\,6866. Panel (a) shows the phase-folded \emph{Kepler} pipeline light curve. Panel (b) represents the amplitude spectrum calculated from the light curve. Panel (c) displays the location of KIC\,8330092 on the CMD of NGC\,6866 marked with a red square. Other orbital variables are marked in black squares.}
    \label{binary-example}
    \end{figure*}

    Our primary focus was identifying stellar pulsations and oscillations, but we also detected binary variability within our region of interest. We identified three targets exhibiting orbital variabilities, such as eclipses or ellipsoidal rotation. One example, KIC\,8330092, is shown in Fig.~\ref{binary-example}. Notably, KIC\,8330092 overlaps with the $\gamma$\,Dor instability strip and is identified as a hierarchical system consisting of an eclipsing close binary system with a third hybrid-pulsating component, as reported by \cite{li2020}.

    \subsection{Surface modulations}

    \begin{figure*}
        \includegraphics[width=\hsize]{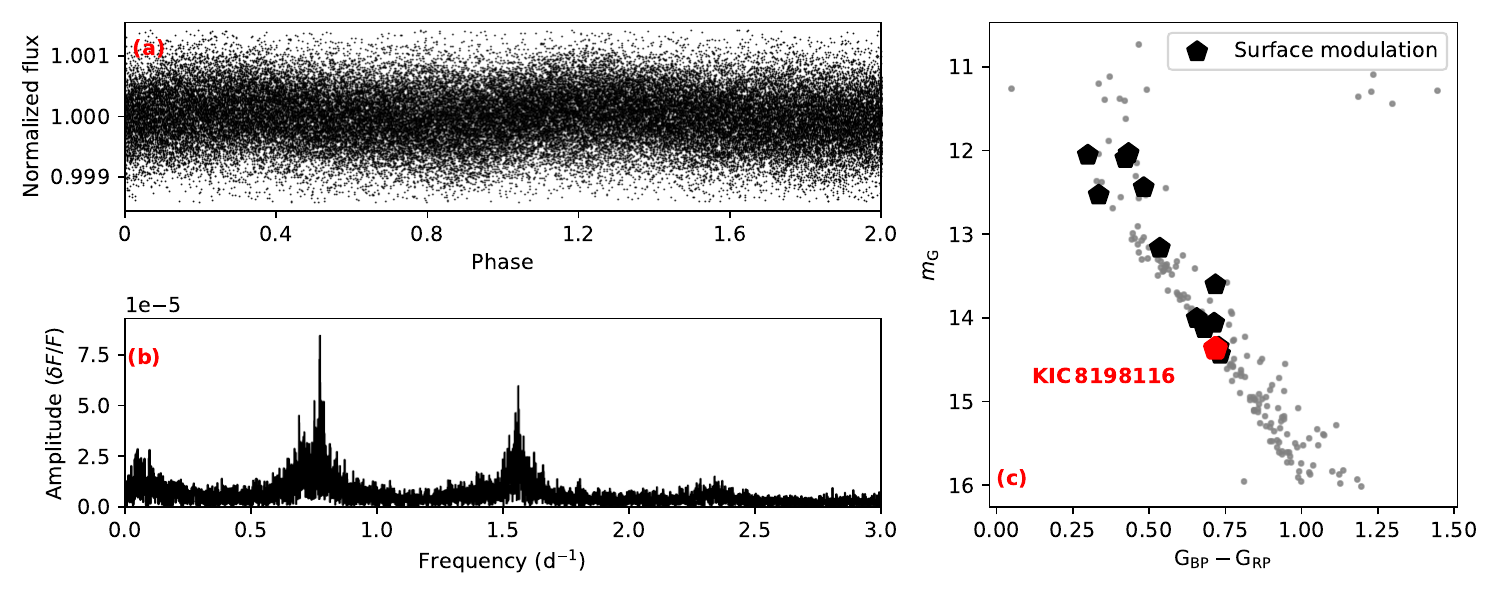}
        \caption{Surface modulation star KIC\,8198116 in NGC\,6866. Panel (a) shows the phase-folded light curve measured from the \emph{Kepler} pipeline light curve. Panel (b) represents the amplitude spectrum calculated from the light curve, zoomed into the low-frequency domain where the profile of the signal and the harmonics can be seen easily. Panel (c) marked the location of KIC\,8198116 on the CMD of NGC\,6866 with a red pentagon. The other surface modulation stars are marked in black pentagons.}
    \label{modulation-example}
    \end{figure*}

    We identified 14 stars exhibiting surface modulation within our region of interest. These stars were characterised by fitting a Lorentzian profile and its harmonics in the low-frequency domain, which indicates the evolution of stellar spots and the non-sinusoidal shape of their light curves. An example of this type of variability is shown in Fig.~\ref{modulation-example}.
\end{appendix}
\end{document}